\documentclass[aps,pra,twocolumn,showpacs]{revtex4}
\usepackage{epsfig,braket}

\newcommand{\up}{\uparrow}
\newcommand{\down}{\downarrow}

\begin{document}

\title{Decoherence and gate performance of coupled solid state qubits}

\author{Markus J.\ Storcz}
\email{storcz@theorie.physik.uni-muenchen.de} \author{Frank K.\
Wilhelm}

\affiliation{Sektion Physik and CeNS,
Ludwig-Maximilians-Universit\"at,  Theresienstr.\ 37, 80333 M\"unchen,
Germany}

\begin{abstract}
Solid state quantum bits are promising candidates for the realization
of a {\em scalable} quantum computer. However, they are usually
strongly limited by decoherence due to the many extra degrees of freedom of
a solid state system.  We investigate a system of two solid
state qubits that are coupled via $\sigma_z^{(i)} \otimes
\sigma_z^{(j)}$ type of coupling. This kind of setup is typical for
{\em pseudospin} solid-state quantum bits such as charge or flux
systems. We evaluate decoherence properties and gate quality factors
in the presence of a common and two uncorrelated baths coupling to
$\sigma_z$, respectively. We show that at low temperatures,
uncorrelated baths do degrade the gate quality more  severely. In
particular, we show that in the case of a common bath, optimum gate
performance of a CPHASE gate can be reached at very low temperatures,
because our type of coupling commutes with the coupling to the
decoherence, which makes this type of coupling attractive as compared
to previously studied proposals with $\sigma_y^{(i)} \otimes \sigma_y^{(j)}$
-coupling. Although less pronounced, this advantage also applies to the CNOT
gate.
\end{abstract}


\maketitle

\section{Introduction}

Quantum computation has been shown to perform certain tasks much
faster than classical computers
\cite{shorsalgorithm,grover,deutsch}. Presently, very mature
physical realizations of this idea originate in atomic physics,
optics, and nuclear magnetic resonance. These systems are phase
coherent in abundance, however, scaling up the existing few-qubits
systems is not straightforward.  Solid state quantum computers have
the potential advantage of being arbitrarily scalable to large systems
of many qubits  \cite{divincenzo1,bkane,orlando}.  Their most
important drawback is the coupling to the many degrees of freedom of a
solid state  system.  Even though recently there has been fast
progress in improving the decoherence  properties of experimentally
realized solid state quantum bits
\cite{nakamura,stonybrook,quantronium,fluxqubitrabi}, this remains a
formidable task.

Quite a lot is known about decoherence properties of single solid
state  qubits, see e.g.\ [\onlinecite{marlies,caspar, makhlin1}], but much
less is known about systems of two or more coupled  qubits
\cite{milburn,governale,thorwart}. However, only for systems of at
least two qubits, the central issue of entanglement can  be
studied. The physically available types of qubit coupling can be
classified as Heisenberg-type exchange which is typical for real
spin-1/2-systems, and Ising-type coupling, which is characteristic for
{\em pseudospin} setups,  where the computational degrees of freedom
are not real spins. In the latter, the different spin components
typically correspond to distinct variables, such as charge and flux
\cite{quantronium,wilhelmmooij} whose couplings can and have to be
engineered on completely different footing.  Previous work
\cite{governale,thorwart} presented the properties of a system  of two
coupled solid state qubits which are coupled via $\sigma_y^{(i)}
\otimes \sigma_y^{(j)}$  type coupling as proposed in \cite{makhlin1}
as the current-current coupling of superconducting charge quantum bits.

On the other hand, many systems such as inductively coupled flux
qubits  \cite{orlando},  capacitively coupled charge qubits
\cite{pashkin,nakamura} and other pseudo-spin  systems \cite{Blick} are
described by a $\sigma_z^{(i)} \otimes \sigma_z^{(j)}$ Ising-type
coupling. This indicates that the computational basis states are coupled,
which i.e.\ in the case
of flux qubits are magnetic fluxes, whereas $\sigma_{x/y}$ are 
electric charges. The $\sigma_z$-observable is a natural way of
coupling, because it is typically easy to couple to. We will study
a two qubit-system coupled this way that
is exposed to Gaussian noise coupling to $\sigma_z$, the ``natural''
observable. This e.g.\ accounts for the crucial effect of
electromagnetic noise in superconducting qubits.  We will compare both
the case of noise that affects both qubits in a correlated way and the
case of uncorrelated  single-qubit errors.  We determine the
decoherence properties of the system by application of the  well known
Bloch-Redfield formalism and determine quality factors of a CNOT gate
for both types of errors and feasible parameters of the system.

\section{Model Hamiltonian}
We model the Hamiltonian of a system of two qubits, coupled via
Ising-type coupling.  Each of the two qubits is a two-state system
that is described in pseudo-spin notation by the single-qubit
Hamiltonian \cite{caspar}
\begin{equation}
\mathbf{H}_{sq} = - \frac{1}{2} \epsilon \hat \sigma_z - \frac{1}{2}
\Delta \hat \sigma_x \textrm{,}
\end{equation}
where $\epsilon$ is the energy bias and $\Delta$ the tunnel matrix
element.  The coupling between the qubits is determined by an extra
term in the Hamiltonian $\mathbf{H}_{\rm qq}=-\frac{K}{2}
\hat{\sigma}_z^{(1)}\otimes  \hat{\sigma}_z^{(2)}$ that represents
e.g.\ inductive interaction (directly or via flux transformer) in the
case of flux qubits \cite{orlando,hannes}. Thus, the complete two-qubit
Hamiltonian in the absence of a dissipative environment reads
\begin{equation} \label{Hop2qb_nobath}
\mathbf{H}_{2qb} = \sum_{i=1,2}\left( -\frac{1}{2} \epsilon_i \hat
\sigma_z^{(i)} - \frac{1}{2} \Delta_i \hat \sigma_x^{(i)} \right) -
\frac{1}{2} K \hat \sigma_z^{(1)} \hat \sigma_z^{(2)}\textrm{.}
\end{equation}
The dissipative (bosonic) environment is conveniently modeled as
either a  common bath or two distinct baths of harmonic oscillators,
coupling to the $\sigma_z$ components of the two  qubits. This
approach universally models baths which produce {\em Gaussian}
fluctuations, such as the noise from linear electrical circuits. An
example for a situation described by a common bath is long correlation
length electromagnetic noise from the experimental environment or
noise  generated or picked up by coupling elements such as flux
transformers  \cite{orlando}. Short correlation length radiation or
local readout and control electronics coupling to individual
qubits \cite{caspar} might be described as coupling to two
uncorrelated baths of harmonic oscillators.

One should note that if the number of qubits is increased to more than
two, there might also occur dissipative effects which neither
affect all qubits nor only a single qubit but rather a cluster of
qubits thus enhancing the complexity of our considerations
\cite{dfsforqc}.

In the case of two uncorrelated baths, the full Hamiltonian reads
\begin{eqnarray} \label{Hop_2baths}
\mathbf{H}_{2qb}^{2b} & = & \sum_{i=1,2}\left( -\frac{1}{2} \epsilon_i
\hat \sigma_z^{(i)} -  \frac{1}{2} \Delta_i \hat \sigma_x^{(i)} +
\frac{1}{2} \hat \sigma_z^{(i)} \widehat{X}^{(i)} \right) {}
\nonumber\\ & &{} - \frac{1}{2} K \hat \sigma_z^{(1)} \hat
\sigma_z^{(2)} + \mathbf{H}_{B_1} + \mathbf{H}_{B_2}\textrm{,}
\end{eqnarray}
where each qubit couples to its own, distinct, harmonic oscillator
bath $\mathbf{H}_{B_i}$, $i=1,2$, via the coupling term
$\sigma_z^{(i)}\widehat{X}^{(i)}$, $i=1,2$,  that bilinearly couples a
qubit to the collective bath  coordinate $\widehat{X}^{(i)}=\zeta
\sum_\nu \lambda_\nu x_\nu$.  We again sum over the two qubits. In the
case of two qubits coupling to one common bath, we model our two qubit
system with the Hamiltonian
\begin{eqnarray} \label{Hop_1bath}
\mathbf{H}_{2qb}^{1b} & = & - \frac{1}{2} \sum_{i=1,2}\left(
\epsilon_i \hat \sigma_z^{(i)} + \Delta_i \hat \sigma_x^{(i)}\right) -
\frac{1}{2} K \hat \sigma_z^{(1)} \hat \sigma_z^{(2)} {} \nonumber\\
& &{} + \frac{1}{2} \left( \hat \sigma_z^{(1)} + \hat \sigma_z^{(2)}
\right) \widehat{X}  + \mathbf{H}_{B}\textrm{,}
\end{eqnarray} 
where $\mathbf{H}_B$ denotes one common bath of harmonic oscillators.

The appropriate starting point for our further analysis is the
singlet/triplet basis, consisting of $\ket{\uparrow\uparrow }:=
(1,0,0,0)^T$, $(1/\sqrt{2})(\ket{\uparrow\downarrow}+\ket{\downarrow
\uparrow}):=(0,1,0,0)^T$, $\ket{\downarrow \downarrow}:=(0,0,1,0)^T$
and $(1/\sqrt{2})(\ket{\uparrow \downarrow}- \ket{\downarrow
\uparrow}):=(0,0,0,1)^T$. In the case of flux qubits, the  $\uparrow$
and $\downarrow$ states correspond to clockwise and counterclockwise
current respectively.

In this basis, the undamped  Hamiltonian $\mathbf{H}_{2qb}$,  equation
(\ref{Hop2qb_nobath}), of the two qubit system assumes the matrix form
\begin{eqnarray} \label{H_2qb}
\mathbf{H}_{2qb} & = &  - \frac{1}{2} \left( \begin{array}{cccc}
  \epsilon+K & \eta & 0 & -\Delta \eta\\ \eta & -K & \eta &
  \Delta\epsilon \\ 0 & \eta & K-\epsilon & \Delta \eta\\ -\Delta \eta
  & \Delta \epsilon & \Delta \eta & -K\\
\end{array} \right)\textrm{,}
\end{eqnarray}
with $\epsilon=\epsilon_1+\epsilon_2$, $\eta=(\Delta_1+\Delta_2)/\sqrt{2}$,
$\Delta \eta = (\Delta_1 - \Delta_2)/\sqrt{2}$ and $\Delta \epsilon = \epsilon_1
- \epsilon_2$.  From now on, we concentrate on the case of equal
parameter settings,  $\Delta_1=\Delta_2$ and $\epsilon_1=\epsilon_2$.

If we now also express the coupling to the  dissipative environment in
this basis, we find in the case coupling to two uncorrelated distinct
baths that
\begin{eqnarray} \label{Hop_matrix_2baths}
\mathbf{H}_{2qb}^{2b} & = &  - \frac{1}{2} \left( \begin{array}{cccc}
  \epsilon-s+K & \eta & 0 & 0\\ \eta & -K & \eta & -\Delta s \\ 0 &
  \eta & K-\epsilon+s & 0\\ 0 & -\Delta s & 0 & -K\\
\end{array} \right) \quad
\end{eqnarray}
with $s=X_1+X_2$ and  $\Delta s = X_1 - X_2$.  Here the bath mediates
transitions between the singlet and triplet states, the singlet is not
a protected subspace.

In the case of two qubits with equal parameters that are coupled to one common
bath we  obtain the matrix
\begin{eqnarray} \label{Hop_matrix_2}
\mathbf{H}_{2qb}^{1b} & = &  - \frac{1}{2} \left( \begin{array}{cccc}
  \epsilon-s+K & \eta & 0 & 0\\ \eta & -K & \eta & 0 \\ 0 & \eta &
  K-\epsilon+s & 0\\ 0 & 0 & 0 & -K\\
\end{array} \right)\textrm{,} \qquad
\end{eqnarray}
where $s=2\widehat{X}$ and $\Delta s = 0$. One directly recognizes
that compared to (\ref{Hop_matrix_2baths}) in this case thermalization
to the singlet state is impeded, because (\ref{Hop_matrix_2}) is
block-diagonal in the singlet and triplet subspaces.  The singlet and
triplet are completely decoupled from each other and in the case of
one common bath the singlet is also completely decoupled from the bath and
thus protected from  dissipative effects. Therefore, a system in
contact with one common bath that is prepared in the singlet state
will never experience any decoherence effects. The singlet state is a
protected subspace (DFS) \cite{whaley}, although a trivial,
one-dimensional one.

\section{Eigenenergies and eigenstates of the two-qubit Hamiltonian}
We calculate exact analytical eigenvalues and eigenvectors of the
unperturbed two-qubit system Hamiltonian in the aforementioned
symmetric case of (\ref{H_2qb}), which reads
\begin{eqnarray} \label{Hop_2qb_ep}
\mathbf{H}_{2qb} & = &  - \frac{1}{2} \left( \begin{array}{cccc}
  \epsilon+K & \eta & 0 & 0\\ \eta & -K & \eta &  0 \\ 0 & \eta &
  K-\epsilon & 0\\ 0 & 0 & 0 & -K\\
\end{array} \right)\textrm{.}
\end{eqnarray}
This Hamiltonian is block-diagonal and the largest block, the triplet,
is three-dimensional, i.e.\ it can be diagonalized using Cardano's
formula.  Details of that calculation are given in Ref.\
\onlinecite{Diplomarbeit}. The case of non-identical qubits is more
easily handled numerically.

\begin{figure}[ht]
\begin{center}
\includegraphics[width=8.5cm]{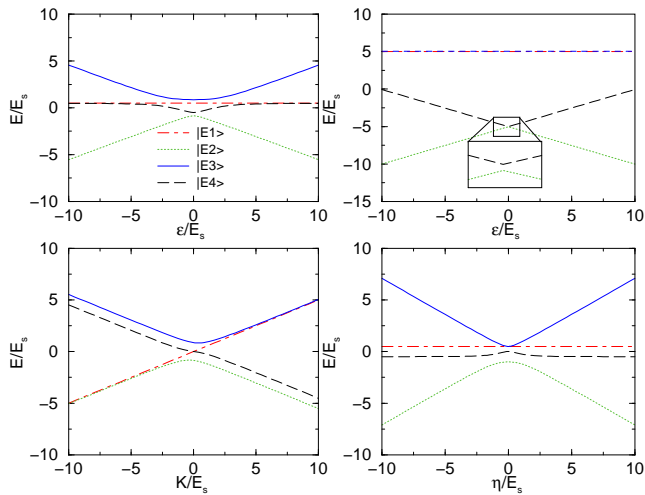}
\end{center}
\caption[Plot of the eigenenergies of the eigenstates $\ket{E1}$,
  $\ket{E2}$, $\ket{E3}$ and $\ket{E4}$.]{Plot of the eigenenergies of
  the eigenstates $\ket{E1}$, $\ket{E2}$, $\ket{E3}$ and
  $\ket{E4}$. From upper left to lower right: 1) $K=\eta=E_s$ and
  $\epsilon$ is varied,  2) $K=10\cdot E_s$, $\eta=E_s$ and $\epsilon$
  is varied; the inset resolves the avoided level crossing due to the
  finite transmission amplitude $\eta$, 3) $\eta=\epsilon=E_s$ and $K$
  is varied, 4) $K=\epsilon=E_s$ and $\eta$ is varied.}
\label{figure_eigenenergies}
\end{figure}

In the following, $\ket{E1}$, $\ket{E2}$, $\ket{E3}$ and $\ket{E4}$
denote the eigenstates of the two-qubit system.  The eigenenergies of
the unperturbed Hamiltonian (\ref{Hop_2qb_ep}) depend on the three
parameters $K$, $\epsilon$ and $\eta$.  Figure
\ref{figure_eigenenergies} displays the eigenenergies in more detail
for typical experimentally accessible values.
The parameters that are choosen for the  parameters $\epsilon$, $\eta$
and $K$ in figure \ref{figure_eigenenergies} correspond to what can be
reached in flux qubits. They typically  assume values  of a few GHz
resembling the parameters of known single- and two-qubit experiments
in Delft \cite{caspar} and at MIT \cite{MIT}.  Therefore, we will
use a characteristic energy scale $E_s$, which is typically $E_s=1$~GHz.
The corresponding scales are
$t_s=1$~ns, $\omega_s=2\pi \cdot 1$~GHz and $T_s=\nu_s (h/k_B)=4.8 \cdot
10^{-2}$~K. Panel 1) shows that
for large values of $\epsilon$, two of the eigenenergies are degenerate
(namely for $\epsilon \gg \eta,K$ the states $\ket{E1}$ and $\ket{E4}$
equal the states $(1/\sqrt{2})(\ket{\up \down}-\ket{\down \up})$ and
$(1/\sqrt{2})(\ket{\up \down}+\ket{\down \up})$, hence the
eigenenergies are degenerate) while near zero energy bias (magnetic
frustation $f=1/2$) all four eigenenergies might be distinguished. Note
also that therefore at zero energy bias the transition frequency
$\omega_{14}=-\omega_{41}$ has a local {\em maximum}, which, as will
be shown below, can only be accessed via non-symmetric driving.

If $K$ is set to a big positive value corresponding to large
ferromagnetic coupling (figure \ref{figure_eigenenergies}, panel 2),
$K=10 \cdot E_s$) the  Hamiltonian (\ref{Hop_2qb_ep}) is nearly
diagonal and hence the eigenstates in good approximation are equal to
the singlet/triplet basis states.  In this case $\ket{E3}$ equals the
triplet state $(1/\sqrt{2})(\ket{\up \down}+\ket{\down \up})$,
$\ket{E2}$ and $\ket{E4}$ equal $\ket{\up \up}$ and $\ket{\down\down}$
respectively for positive values of $\epsilon$. For large negative
values of $\epsilon$, the two states $\ket{E2}$ and $\ket{E4}$ become
equal $\ket{\down \down}$ and $\ket{\up \up}$ with a pseudo-spin-flip
between clockwise and counter-clockwise rotating current at
$\epsilon=0$ when going from positive to negative $\epsilon$.
In the case of large ferromagnetic coupling, the
ground state tends towards the superposition $(1/\sqrt{2})(\ket{\up
\up} + \ket{\down \down})$. Panel 2) shows that only for
$\epsilon$ equal to zero, both $\ket{E2}=\ket{\up \up}$
($\ket{E2}=\ket{\down\down}$ for negative $\epsilon$) and
$\ket{E4}=\ket{\down \down}$ ($\ket{E4}=\ket{\up\up}$ for negative
$\epsilon$) have the same energies (which one would expect if the
$-(1/2)K\sigma_z^{(1)}\sigma_z^{(2)}$ term in the Hamiltonian
dominates), because if $\epsilon$ is increased the $\epsilon_i \hat
\sigma_z^{(i)}$ (i=1,2) terms  in the Hamiltonian change the energy.

For large antiferromagnetic coupling, $\vert -K \vert \gg \epsilon,
\Delta$  the  states $\ket{\up \down}$ and $\ket{\down \up}$ are
favorable.  In this limit, the ground state tends towards
$(1/\sqrt{2})(\ket{\up\down}+\ket{\down\up})$ and the energy splitting
between $(1/\sqrt{2})(\ket{\up \down} + \ket{\down \up})$ and
$(1/\sqrt{2})(\ket{\up \down}-\ket{\down \up})$ vanishes
asymptotically,  leaving the ground state nearly degenerate.

\begin{figure}[t]
\begin{center}
\includegraphics[width=8.5cm]{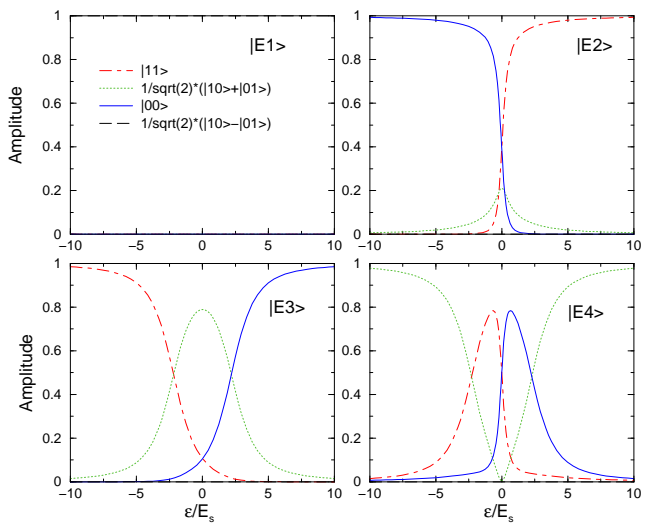}
\end{center}
\caption[Plot of the amplitude of the different singlet/triplet states
of which the eigenstates denoted by $\ket{E1}$, $\ket{E2}$, $\ket{E3}$
and $\ket{E4}$ are composed for the four eigenstates.]{Plot of the
amplitude of the different singlet/triplet states of which the
eigenstates denoted by $\ket{E1}$, $\ket{E2}$, $\ket{E3}$ and
$\ket{E4}$ are composed for the four eigenstates. In all plots
$\epsilon$ is varied and $K$ and $\eta$ are fixed to $E_s$.}
\label{figure_state_amplitudes}
\end{figure}
From figure \ref{figure_eigenenergies}, panel 3), one directly
recognizes that the singlet eigenenergy crosses the triplet spectrum,
which is a consequence of the fact  that the singlet does not interact
with any triplet states.  At zero energy bias (magnetic frustration
$f=1/2$ for a flux qubit),  none of the eigenstates equal one of the
triplet basis states  (e.g.\ as observed for a large energy bias
$\epsilon$),  they are rather nontrivial superpositions.  This is
elucidated further in the next paragraph.  The inset of panel 2)
depicts the level anti-crossing between the eigenenergies of the two
states $\ket{E2}$ and $\ket{E4}$ due to quantum tunneling.

In general, the eigenstates are a superposition of singlet/triplet
states.  Figure \ref{figure_state_amplitudes} shows, how
singlet/triplet states combine into eigenstates for different qubit
parameters. The first eigenstate $\ket{E1}$ equals $(1/\sqrt{2})(\ket{\up \down}
-\ket{\down \up})$ for all times while the other eigenstates $\ket{E2}$,
$\ket{E3}$ and $\ket{E4}$ are in general superpositions of the
singlet/triplet basis states.  For large values of $\vert \epsilon
\vert$, the eigenstates approach the singlet/triplet basis states. In
particular  at typical working points, where $\epsilon \approx 5 \cdot
\Delta$ \cite{caspar} the eigenstates already nearly equal the
singlet/triplet basis states. Hence, although the anticrossing
described above corresponds to the anticrossing used in
\cite{stonybrook,casparscience} to demonstrate Schr\"odinger's cat
states, {\em entanglement} is prevalent away from the degeneracy
point. For an experimental proof, one still would have to show that
one has successfully prepared coherent couplings by  spectroscopically
tracing the energy spectrum.  Note, that for clarity,  in figure
\ref{figure_state_amplitudes} the inter-qubit coupling strength $K$ is
fixed to a rather high value of $E_s$ that also sets the width of the
anti-crossing, which potentially can be very narrow.

\begin{figure}[t]
\begin{center}
\includegraphics[width=8.5cm]{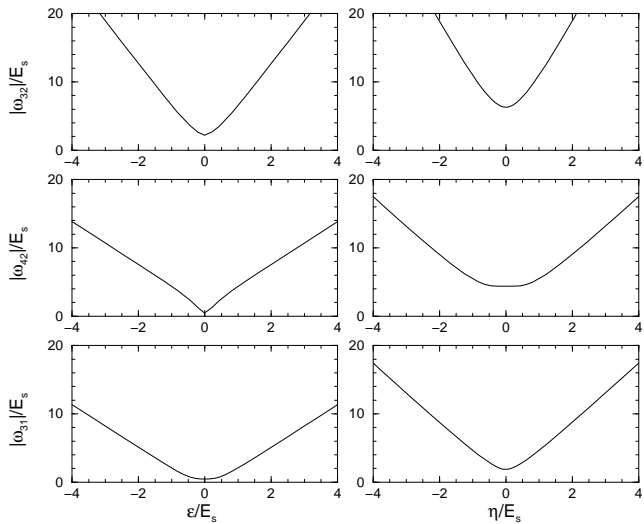}
\end{center}
\caption[Plot of the transition frequencies $\omega_{32}$,
$\omega_{42}$ and $\omega_{31}$.]{Plot of the absolute value of the
transition frequencies $\omega_{32}$, $\omega_{42}$ and
$\omega_{31}$. In the left column $K=\eta=0.2\cdot E_s$ and $\epsilon$
is varied. Right column $K=0.2\cdot E_s$, $\epsilon=E_s$ and $\eta$ is
varied.}
\label{omegas1}
\end{figure}
\begin{figure}[t]
\begin{center}
\includegraphics[width=8.5cm]{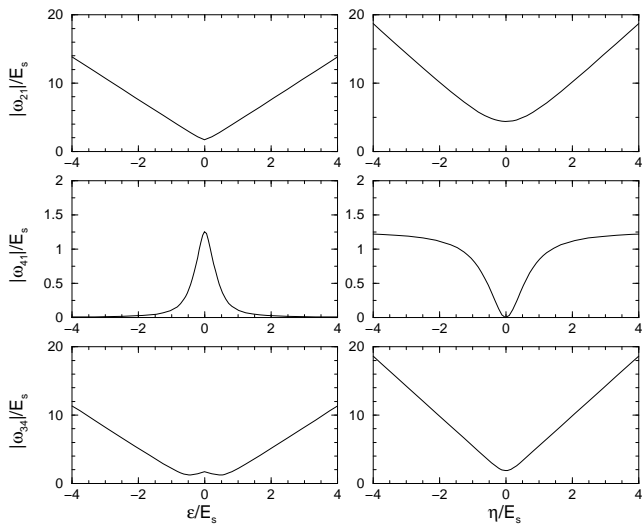}
\end{center}
\caption[Plot of the transition frequencies $\omega_{21}$,
$\omega_{41}$ and $\omega_{34}$.]{Plot of the transition frequencies
$\omega_{21}$, $\omega_{41}$ and $\omega_{34}$. In the left column
$K=\eta=0.2 \cdot E_s$ and $\epsilon$ is varied. Right column
$K=0.2\cdot E_s$, $\epsilon=E_s$ and $\eta$ is varied.}
\label{omegas2}
\end{figure}

\subsection{Spectroscopy}
As a first technological step towards demonstrating coherent
manipulation of qubits, usually the transition frequencies between
certain energy levels are probed \cite{casparscience,stonybrook},
i.e.\ the energy {\em differences} between the levels.  Figures
\ref{omegas1} and \ref{omegas2} depict the transition frequencies
between the four eigenstates. The transition frequencies are defined
as $\omega_{nm}=(E_n-E_m)/\hbar$ and $\omega_{nm}=-\omega_{mn}$. The
transitions between the singlet state $\ket{E1}$ and the triplet
states are forbidden in the case of one common bath, due to the
special symmetries of the Hamiltonian (\ref{Hop_1bath}), if the system
is driven collectively through a time dependent energy bias
$\epsilon_1(t)= \epsilon_2(t)$.  However, in the case of two distinct
baths the environment can mediate transitions between the singlet
state and the triplet states.

Not all transition frequencies have local minima at $\epsilon=0$.  The
frequencies $\omega_{41}$ and $\omega_{34}$ have local maxima at zero
energy bias $\epsilon$. This can already be inferred from  figure
\ref{figure_eigenenergies}, panel 1), the energy of the eigenstate
$\ket{E4}$ has a local minimum at $\epsilon=0$. Similarily, the
substructure of $\omega_{34}$ can be understood from figure
\ref{figure_eigenenergies}: the frequency $\omega_{34}$ has a local
maximum at $\epsilon=0$, because of the local minimum of the
eigenenergy of the state $\ket{E4}$. First, if $\epsilon$ is
increased, the level spacing of $\ket{E4}$ and $\ket{E3}$
decreases. Then, for larger values of $\epsilon$ the level spacing of
$\ket{E4}$ and $\ket{E3}$ increases again. Thus, the structure
observed for $\omega_{34}$ around $\epsilon=0$ emerges in figure
\ref{omegas2}.

\section{Bloch-Redfield-Formalism} \label{ch_bloch_redfield_formalism}
In order to describe decoherence in the weak damping limit, we use the
Bloch-Redfield-Formalism \cite{argyres}. It provides a systematic way
of finding a set of coupled master equations which describes the
dynamics of the reduced (i.e.\ the reservoir coordinates are traced
out) density matrix  for a given system in contact with a dissipative
environment and has recently been shown to be numerically  equivalent
to the more elaborate path-integral scheme \cite{Hartmann}.  The
Hamiltonian of our two qubit system in contact with a dissipative
environment, eqs. (\ref{Hop_2baths}) and (\ref{Hop_1bath}), has the generic
``system+bath'' form
\begin{equation}
\mathbf{H}_{op}(t) =
\mathbf{H}_{2qb}+\mathbf{H}_B+\mathbf{H}_{int}\textrm{,}
\end{equation}
where $\mathbf{H}_B$ is a bath of harmonic oscillators and
$\mathbf{H}_{int}$ inherits the coupling to a dissipative
environment. In our case the effects of driving are not investigated.
In Born approximation and when the system is only weakly coupled to
the environment, Bloch-Redfield theory provides the following set of
equations for the reduced density matrix $\rho$ describing the
dynamics of the system \cite{blum,weiss}
\begin{equation} \label{redfield_eqns}
\dot \rho_{nm}(t) = -i\omega_{nm} \rho_{nm}(t) - \sum_{kl} R_{nmk\ell}
\rho_{k \ell}(t) \textrm{,}
\end{equation}
where $\omega_{nm}=(E_n-E_m)/\hbar$, and $\max_{n,m,k, \ell} \limits
\vert \mbox{Re}(R_{nmk \ell}) \vert < \min_{n \neq m} \limits \vert
\omega_{nm} \vert$ must hold. The Redfield relaxation tensor
$R_{nmk\ell}$  comprises the dissipative effects of the coupling of
the system to the environment.  The elements of the Redfield
relaxation tensor are given through golden rule rates \cite{weiss}
\begin{eqnarray} \label{redfield_tensor}
R_{nmk\ell} & = & \delta_{\ell m}\sum_{r}\Gamma_{nrrk}^{(+)}+
\delta_{nk}\sum_{r}\Gamma_{\ell rrm}^{(-)} {} \nonumber\\ & &{}
-\Gamma_{\ell mnk}^{(+)}-\Gamma_{\ell mnk}^{(-)}\textbf{.}
\end{eqnarray}
 
\subsection{Two qubits coupled to two distinct baths}
We now evaluate the Golden rule expressions in (\ref{redfield_tensor})
in the case of two qubits, each coupled to a distinct harmonic
oscillator bath. Here,
$\widetilde{H}_I(t)=\exp(iH_Bt/\hbar)H_I\exp(-iH_Bt/\hbar)$ denotes
the coupling between system and bath in the interaction picture, and
the bracket denotes thermal average of the bath degrees of
freedom. Writing down all contributions gives
\begin{eqnarray}
\Gamma_{\ell mnk}^{(+)} & = & \hbar^{-2} \int_0^\infty dt \
	e^{-i\omega_{nk}t} \bigl< e^{i(H_{B_{1}}+H_{B_{2}})t/\hbar)}
	{} \nonumber \\ & &{} \times \Bigl(\sigma_{z,\ell
	m}^{(1)} \otimes \widehat{X}^{(1)} + 
	\sigma_{z,\ell m}^{(2)} \otimes \widehat{X}^{(2)}\Bigr) {} \nonumber \\
	& &{} \times
	e^{-i(H_{B_1}+H_{B_2})t/\hbar)} \nonumber \\ & &{}
	\times \Bigl(\sigma_{z,nk}^{(1)}  \otimes
	\widehat{X}^{(1)}+ \sigma_{z,nk}^{(2)}  \otimes
	\widehat{X}^{(2)} \Bigr) \bigr> \textrm{,}
\end{eqnarray}
where $\sigma_{z,nm}^{(i)}$ ($i=1,2$) are the matrix elements of $\hat
\sigma_z^{(i)}$ with respect to the eigenbasis of the unperturbed
Hamiltonian (\ref{Hop_2qb_ep}) and likewise for $\Gamma_{\ell m nk}^{(-)}$.

We assume ohmic spectral densities with a Drude-cutoff. This is a
realistic assumption i.e.\ for electromagnetic noise \cite{caspar} and
leads to integrals in the rates which are tractable by the residue
theorem.  The cutoff frequency $\omega_c$ for the spectral functions
of the two qubits is typically assumed to be  the largest frequency in
the problem, this is discussed further in section
\ref{renormalization_effects}
\begin{equation} \label{spectral_functions_2b}
J_1(\omega) = \frac{\alpha_1 \hbar
\omega}{1+\frac{\omega^2}{\omega_c^2}}\textrm{ and } J_2(\omega) =
\frac{\alpha_2 \hbar \omega}{1+\frac{\omega^2}{\omega_c^2}}\textrm{.}
\end{equation}
The dimensionless parameter $\alpha$ describes the strength of the
dissipative effects that enter the Hamiltonian via the coupling to the
environment, described by $s$ and $\Delta s$. In order for the
Bloch-Redfield formalism, which involves  a Born approximation in the
system-bath coupling, to be valid,  we have to assume $\alpha_{\rm
1/2}\ll1$.  After tracing out over the bath degrees of freedom, the
rates read
\begin{eqnarray}
\Gamma_{\ell m n k}^{(+)} & = & \frac{1}{8\hbar} \left[ \Lambda^1
			J_1(\omega_{nk})+ \Lambda^2 J_2(\omega_{nk})
			\right] {} \nonumber \\ & & \times
			\left( \coth (\beta \hbar \omega_{nk}/2)-1
			\right) {} \nonumber \\ & &{} +
			\frac{i}{4\pi \hbar} \Bigg [ \Lambda^2
			M(\omega_{nk},2) + \Lambda^1 M(\omega_{nk},1)
			\Bigg ]
			\label{gammap_2b}
\end{eqnarray}
with $\Lambda^1 = \Lambda^1_{\ell m n k} = \sigma_{z,\ell m}^{(1)} \sigma_{z,nk}^{(1)}$,
$\Lambda^2 = \Lambda^2_{\ell m n k} =  \sigma_{z,lm}^{(2)} \sigma_{z,nk}^{(2)}$ and
\begin{equation}
M(\Omega,i) = {\cal P} \int_0^\infty \limits d \omega \
\frac{J_i(\omega)}{\omega^2-\Omega^2} (\coth(\beta \hbar
\omega/2)\Omega + \omega)\textrm{,}
\end{equation}
here ${\cal P}$ denotes the principal value. Likewise,
\begin{eqnarray}
\Gamma_{\ell m n k}^{(-)} & = & \frac{1}{8\hbar} \left[ \Lambda^1
			J_1(\omega_{\ell m})+ \Lambda^2
			J_2(\omega_{\ell m}) \right] {}
			\nonumber \\ & & \times \left( \coth (\beta
			\hbar \omega_{\ell m}/2)+1 \right) {}
			\nonumber \\ & &{} + \frac{i}{4\pi \hbar}
			\Bigg [ \Lambda^2 M(\omega_{lm},2) + \Lambda^1
			M(\omega_{lm},1) \Bigg ]\textrm{.} \quad
			\label{gammam_2b} \label{eq:rate_2bath}
\end{eqnarray}
The rates $\Gamma_{\ell m n k}^{(+)}$ and $\Gamma_{\ell m n k}^{(-)}$
might be inserted into (\ref{redfield_tensor}) to build the Redfield
tensor. Note here that for $\omega_{nk} \rightarrow 0$ and
$\omega_{lm} \rightarrow 0$ respectively, the real part of the rates
(which is responsible for relaxation and dephasing) is of value
$\Gamma_{\ell m n k}^{(+)} = \Gamma_{\ell m n k}^{(-)} =
\frac{1}{4\beta \hbar} \left[ \sigma_{z,\ell m}^{(1)}
\sigma_{z,nk}^{(1)} \alpha_1 + \sigma_{z,lm}^{(2)} \sigma_{z,nk}^{(2)}
\alpha_2 \right]$.

To solve the set of differential equations  (\ref{redfield_eqns}), it
is convenient not to use the superoperator notation
but either to write $\rho$ as a vector.
In general the Redfield equations (\ref{redfield_eqns}) without
driving are solved by an ansatz of the type $\rho(t)=B \exp(\tilde
R_i) B^{-1} \rho(0)$, where $\tilde R_i$ is a diagonal matrix. The
entries of this diagonal matrix are the eigenvalues of the Redfield
tensor (\ref{redfield_tensor}), written in matrix form, including the
dominating term $i\omega_{nm}$ (cf. equation \ref{redfield_eqns}).
Here, the reduced density matrix $\rho=(\rho_{11},\ldots,
\rho_{44})^T$ is written as a vector.  The matrix $B$ describes the
basis change to the eigenbasis of $\tilde R_i$, in which $\tilde R_i$
has diagonal form.

\subsection{Two qubits coupled to one common bath}
For the case of two qubits coupled to one common bath we perform the
same calculation as in the last section, which leads to expressions
for the rates analogous to eqs.\ (\ref{eq:rate_2bath})
\begin{eqnarray}
\Gamma_{\ell m n k}^{(+)} & = & \frac{1}{8\hbar} \Lambda
		J(\omega_{nk}) \left( \coth (\beta  \hbar
		\omega_{nk}/2)-1 \right) + \frac{i\Lambda}{4\pi \hbar}
		{} \nonumber \\  & &\times {\cal P}
		\int_0^\infty \limits d\omega \ \frac{J(\omega)}
		{\omega^2-\omega_{nk}^2} \left( \coth (\beta \hbar
		\omega/2)\omega_{nk}-\omega \right)\textrm{,}\nonumber
		\\
\end{eqnarray}
with $\Lambda= \Lambda_{\ell m n k}=\sigma_{z,\ell m}^{(1)} \sigma_{z,nk}^{(1)}
		+\sigma_{z,\ell m}^{(1)} \sigma_{z,nk}^{(2)} +
		\sigma_{z,\ell m}^{(2)} \sigma_{z,nk}^{(1)} +
		\sigma_{z,lm}^{(2)}\sigma_{z,nk}^{(2)}$,  and
\begin{eqnarray}
\Gamma_{\ell m n k}^{(-)} & = & \frac{1}{8\hbar} \Lambda
		J(\omega_{\ell m}) \left( \coth (\beta  \hbar
		\omega_{\ell m}/2)+1 \right)  + \frac{i \Lambda}{4\pi
		\hbar} \nonumber\\ & &\times {\cal P}
		\int_0^\infty \limits d\omega \ \frac{J(\omega)}
		{\omega^2-\omega_{\ell m}^2} \left( \coth (\beta \hbar
		\omega/2)\omega_{\ell m} +\omega
		\right)\textrm{,}\nonumber \\
\label{eq:rate_1bath}
\end{eqnarray}
The difference between the rates for the case of two distinct baths
(\ref{gammap_2b}), (\ref{gammam_2b}) are the two extra terms
$\sigma_{z,lm}^{(1)} \sigma_{z,nk}^{(2)}$ and  $\sigma_{z,lm}^{(2)}
\sigma_{z,nk}^{(1)}$.  They originate when tracing out the bath
degrees of freedom. In the case of one common bath there is only one
spectral function, which we also assume to be ohmic $J(\omega) =
(\alpha \hbar \omega)/(1+\frac{\omega^2}{\omega_c^2})$. For
$\omega_{nk} \rightarrow 0$ and $\omega_{lm} \rightarrow 0$
respectively the real part of the rates is of the value $\Gamma_{\ell
m n k}^{(+)} = \Gamma_{\ell m n k}^{(-)} =  \frac{\alpha}{4\beta
\hbar} \Lambda \textrm{, for }\omega_{lm},\omega_{nk}\rightarrow 0$.

\subsection{Dynamics of coupled flux qubits with dissipation}
The dissipative effects affecting the two qubit system lead to
decoherence, which manifests itself in two ways: The system
experiences energy relaxation on a time scale $\tau_R=\Gamma_R^{-1}$
($\Gamma_R$ is the sum of the relaxation rates of the four diagonal
elements of the reduced density matrix; $\Gamma_R=-\sum_n \Lambda_n$
and $\Lambda_n$ are the eigenvalues of the matrix $R_{n,m,n,m}$,
$n,m=1,\ldots , 4$.), called relaxation time, into a thermal mixture of
the system's energy eigenstates. Therefore the diagonal elements of
the reduced density matrix decay to the value given by the Boltzmann
factors.  The quantum coherent dynamics of the system are superimposed
on the relaxation and decay on a usually shorter time scale
$\tau_{\varphi_{ij}}=\Gamma_{\varphi_{ij}}^{-1}$ ($i,j=1,\ldots,4;
i\neq j$ and $\Gamma_{\varphi_{nm}}=-\mbox{Re}R_{n,m,n,m}^{1b,2b}$)
termed dephasing time. Thus dephasing causes the off diagonal terms
(coherences) of the reduced density matrix to tend towards zero.

\begin{figure}[t]
\begin{center}
\includegraphics[width=8.5cm]{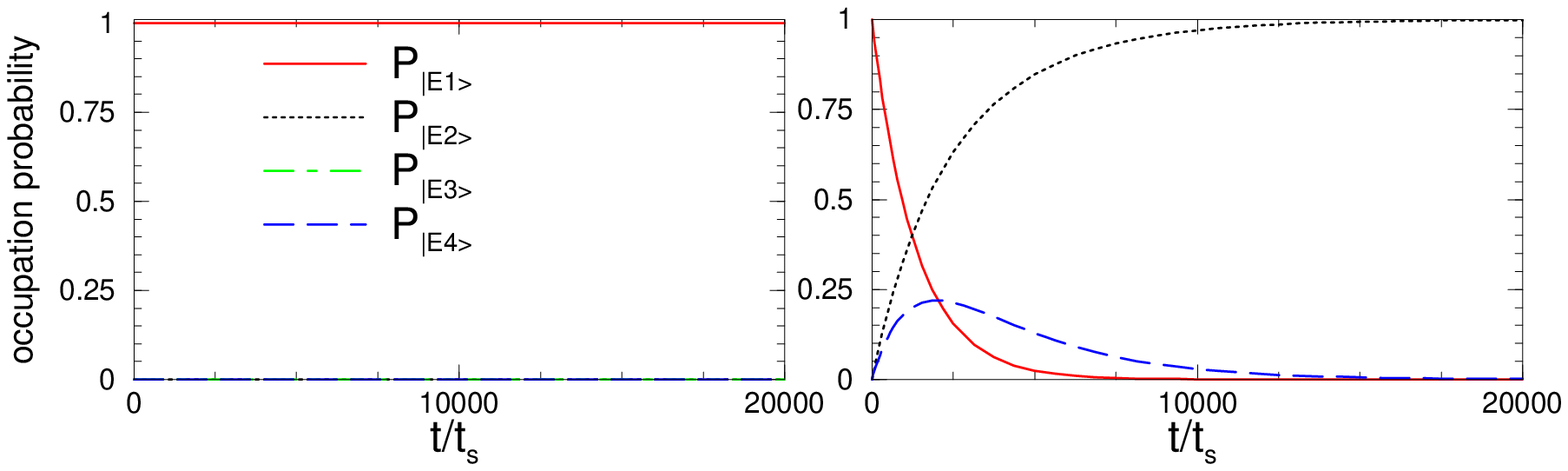}
\includegraphics[width=8.5cm]{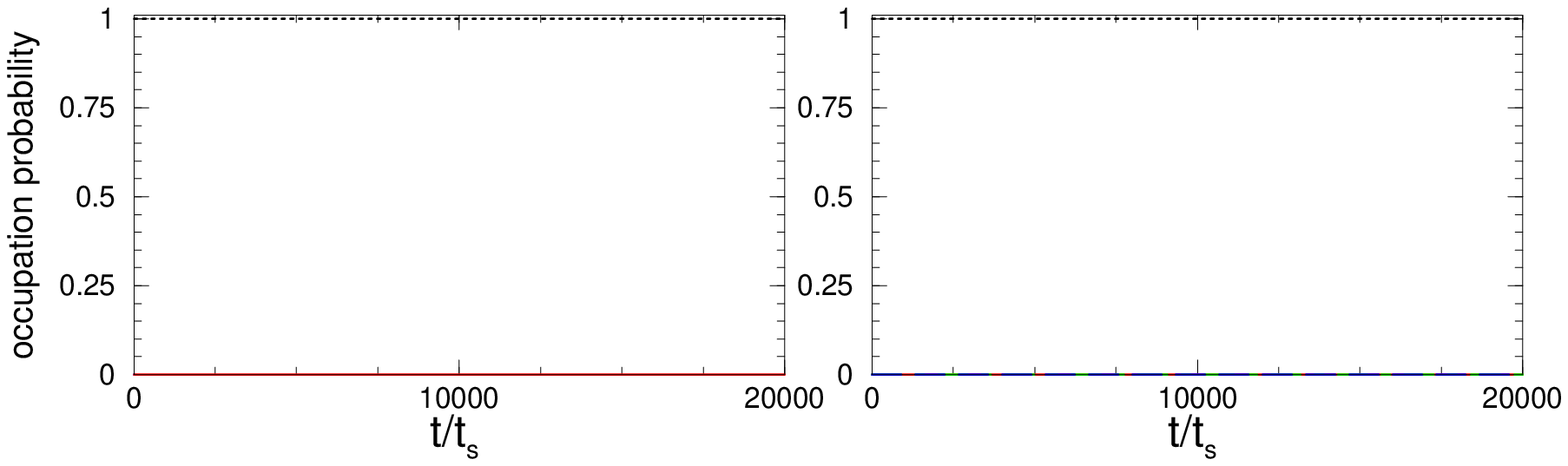}
\includegraphics[width=8.5cm]{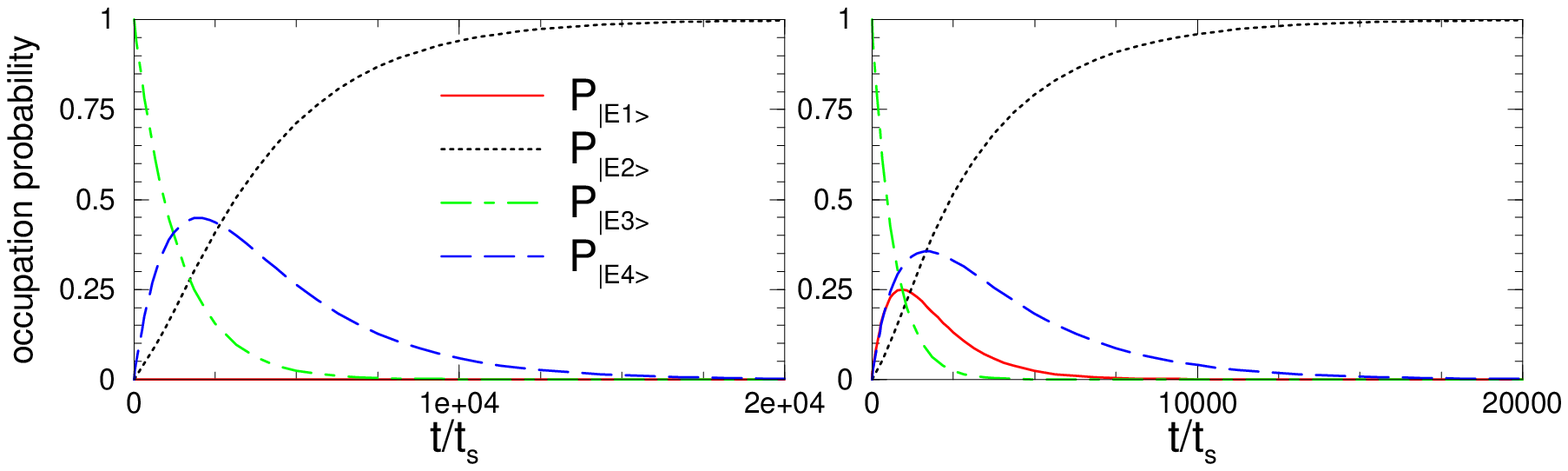}
\end{center}
\caption[Plot of the occupation probability of the four eigenstates
$\ket{E1},\ket{E2}, \ket{E3}$ and $\ket{E4}$ for initially starting in
one of the eigenstates at $T=0$~K.]{Plot of the occupation probability
of the four eigenstates $\ket{E1},\ket{E2}, \ket{E3}$ and $\ket{E4}$
for initially starting in one of the eigenstates $\ket{E1}$ (first row),
$\ket{E2}$ (second row) or $\ket{E3}$ (third row) at $T=0$~K.  The left
column illustrates the case of two qubits coupling to one common bath
and the right column the case of two qubits coupling to two distinct
baths. The energies $K$, $\epsilon$ and $\eta$ are all fixed to
$E_s$. The characteristic timescale $t_s$ is $t_s=1/E_s$.}
\label{fig_relax_1}
\end{figure}

\begin{figure}[t]
\begin{center}
\includegraphics[width=8.5cm]{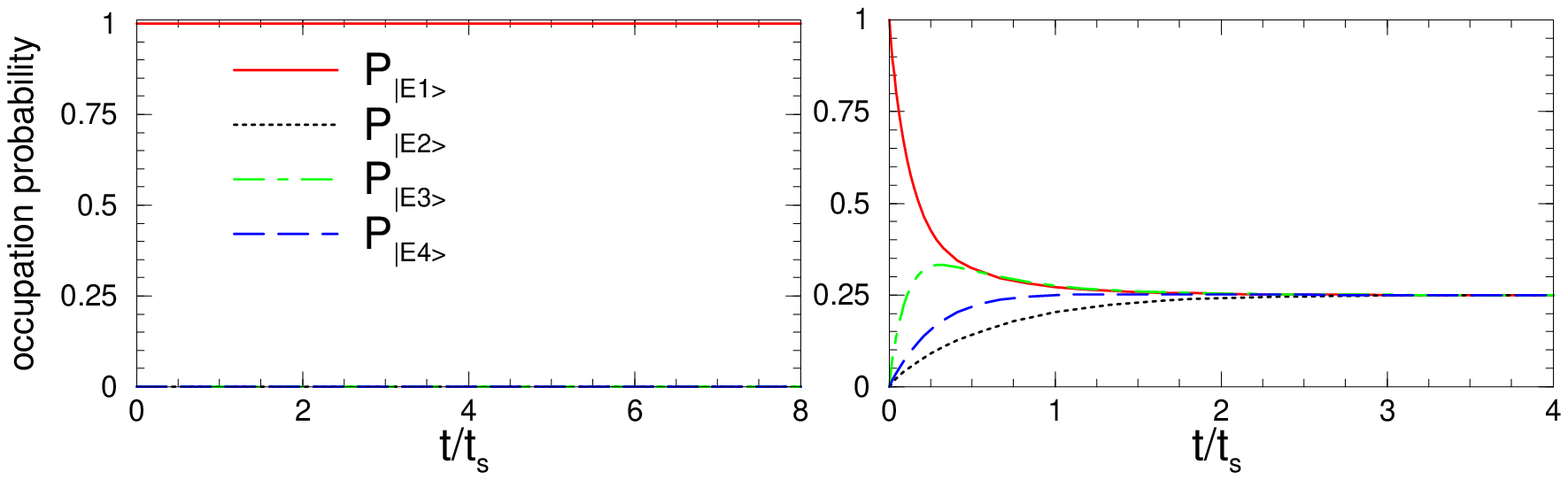}
\includegraphics[width=8.5cm]{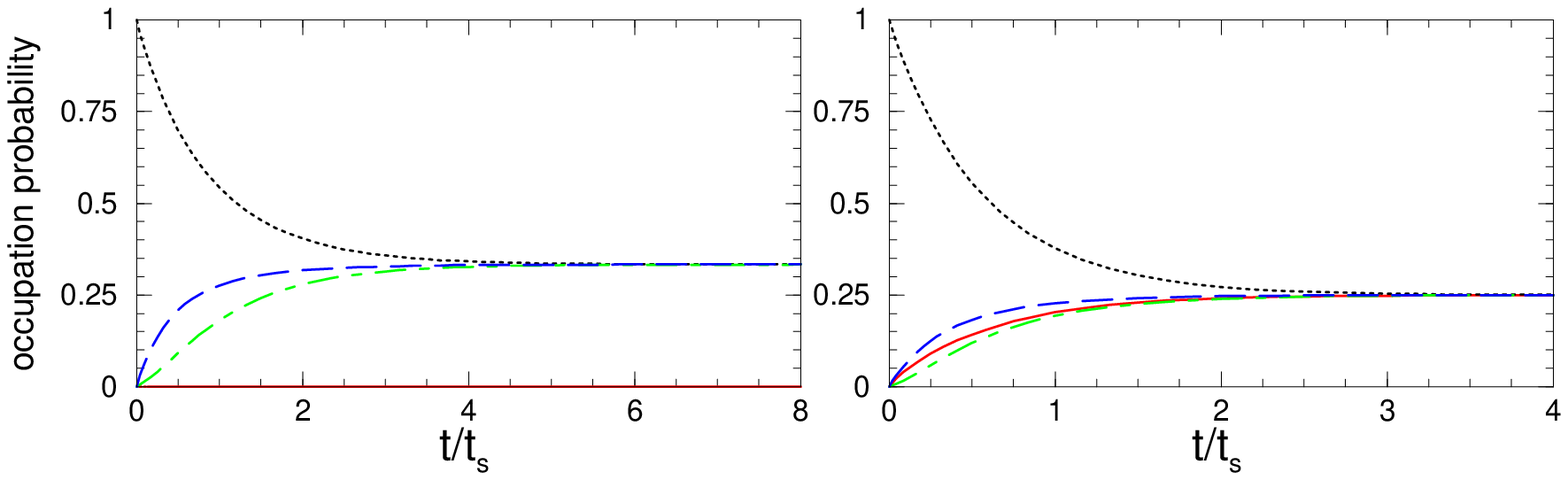}
\end{center}
\caption[Plot of the occupation probability of the four eigenstates
$\ket{E1},\ket{E2}, \ket{E3}$ and $\ket{E4}$ for initially starting in
one of the eigenstates at $T=21E_s$.]{Plot of the occupation
probability of the four eigenstates $\ket{E1},\ket{E2}, \ket{E3}$ and
$\ket{E4}$ for initially starting in one of the eigenstates $\ket{E1}$
(upper row) or $\ket{E2}$ (lower row) at
$T=21 \cdot E_s$.  The left column illustrates the case of two qubits
coupling to one common bath and the right column the case of two
qubits coupling to two distinct baths. The energies $K$, $\epsilon$
and $\eta$ are all fixed to $E_s$. The characteristic timescale $t_s$
is $t_s=1/E_s$.}
\label{fig_relax_2}
\end{figure}

First, we investigate the incoherent relaxation of the two qubit
system out of  an eigenstate.  At long times, the system is expected
to reach  thermal equilibrium, $\rho_{eq}=(1/Z)e^{-\beta H}$. Special
cases are $T=0$ where $\rho_{eq}$ equals the projector on the ground
state and $T \rightarrow \infty$ where all eigenstates are occupied
with the same probability, i.e.\ $\rho_{\rm eq}=(1/4)\hat{1}$.  
Figure \ref{fig_relax_1} and
\ref{fig_relax_2} illustrate the relaxation of the system prepared in
one of the four eigenstates for temperatures $T=0$ and  $T=21\cdot
E_s$ respectively. The qubit energies $K$, $\epsilon$ and $\eta$ are
all set to $E_s$ and $\alpha$ is set to $\alpha=10^{-3}$.  From figure
\ref{figure_eigenenergies} one recognizes relaxation into the
eigenstate $\ket{E2}$,  the ground state for this set of parameters.

At low temperatures ($T=0$), we observe that for the case of two
distinct uncorrelated baths a system prepared in one of the four
eigenstates always relaxes into the ground state. In the case of two
qubits coupling to one common bath, this is not always the case, as can
be seen in the upper left panels of figs.\ \ref{fig_relax_1} and
\ref{fig_relax_2}. This can be explained through our previous
observation, that the singlet is a protected subspace: Neither the
free nor, unlike in the case of distinct baths, the  bath-mediated
dynamics couple the singlet to the triplet space.  Moreover, we can
observe that relaxation to the ground state happens by populating
intermediate eigenstates with a lower energy than the initial state
the system was prepared in at $t=0$ (cf. figure
\ref{figure_eigenenergies}).

For high temperatures ($T\simeq 21\cdot E_s$) the system thermalizes
into thermal equilibrium, where all eigenstates have equal occupation
probabilities. Again, in the case of one common bath, thermalization of
the singlet state is impeded  and  the three eigenstates
$\ket{E2}$, $\ket{E3}$ and $\ket{E4}$ have equal occupation
probabilites of $1/3$ after the relaxation time.

\begin{figure}[t]
\begin{center}
\includegraphics[width=8.5cm]{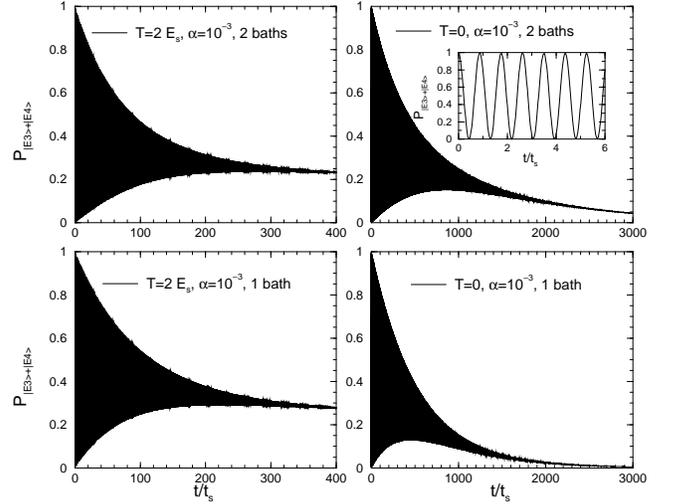}
\end{center}
\caption[Plot of the occupation probability
$P_{(1/\sqrt{2})(\ket{E3}+\ket{E4})}(t)$ when starting in the initial
state $(1/\sqrt{2})(\ket{E3}+\ket{E4})$ at $T=0.1$ K.]{Plot of the
occupation probability $P_{(1/\sqrt{2})(\ket{E3}+\ket{E4})}(t)$ when
starting in the initial state $(1/\sqrt{2})(\ket{E3}+\ket{E4})$, which
is a superposition of eigenstates $\ket{E3}$ and $\ket{E4}$. First row
shows the behaviour for two qubits coupling to two uncorrelated
baths. The lower row shows the behaviour for two qubits coupled to one
common bath. The qubit parameters $\epsilon$, $\eta$ and $K$ are set
to $E_s$, $\alpha$ is set to $\alpha=10^{-3}$. The inset resolves the
time scale of the coherent oscillations.}
\label{fig_dynamics_2}
\end{figure}

If the system is prepared in a superposition of eigenstates, e.g.\
$\ket{E3}$ and $\ket{E4}$ as in Fig.\ \ref{fig_dynamics_2}, which are
not in a protected subspace,  we observe coherent oscillations between
the eigenstates that are damped due to dephasing and after the
decoherence time the occupation probability of the eigenstates is
given by the Boltzmann factors. This behaviour is depicted in figure
\ref{fig_dynamics_2}. Here for $\alpha=10^{-3}$, the cases of $T=0$ and
$T=2.1 \cdot E_s$ are compared. When the temperature is low enough, the
system will relax into the ground state $\ket{E2}$, as illustrated by the right
column of figure \ref{fig_dynamics_2}. Thus the occupation probability
of the state $(1/\sqrt{2})(\ket{E3}+\ket{E4})$ goes to zero. Here, in the case
of zero temperature,
the decoherence times for the case of one common or two distinct baths
are of the same order of magnitude. The left column illustrates the
behaviour when the temperature is increased. At $T=2.1 \cdot E_s$ the
system relaxes into an equally populated state on times much shorter
than for $T=0$.
For low temperatures the characteristic  time scale for dephasing and
relaxation is somewhat shorter for the case of one common bath
($\tau^{1b}/\tau^{2b} \approx 0.9$ for
$\alpha=10^{-3}$). This can be explained by observing the temperature
dependence of the rates shown in figure
\ref{fig_temp_dependence_1}. Though for the case of one common bath
two of the dephasing rates are zero at $T=0$, the remaining rates are
always slightly bigger for the case of one common bath compared to the
case of two distinct baths. If the system is prepared in a  general
superposition, here $\ket{E3}$ and $\ket{E4}$, nearly all rates become
important thus compensating the effect of the two rates which are
approximately zero at zero temperature and leading to faster
decoherence.

If $\alpha$ and therefore the strength of the dissipative effects is
increased from $\alpha=10^{-3}$ to $\alpha=10^{-2}$, the observed
coherent motion is significantly damped. Variation of $\alpha$ leads
to a phase shift of the coherent oscillations, due to renormalization
of the frequencies \cite{governale}.  However, in our case the effects
of renormalization are very small, as discussed in chapter
\ref{renormalization_effects}, and cannot be observed in our plots.

\subsection{Temperature dependence of the rates} \label{temp_dependence_rates}
\begin{figure}[t]
\begin{center}
\includegraphics[width=8.5cm]{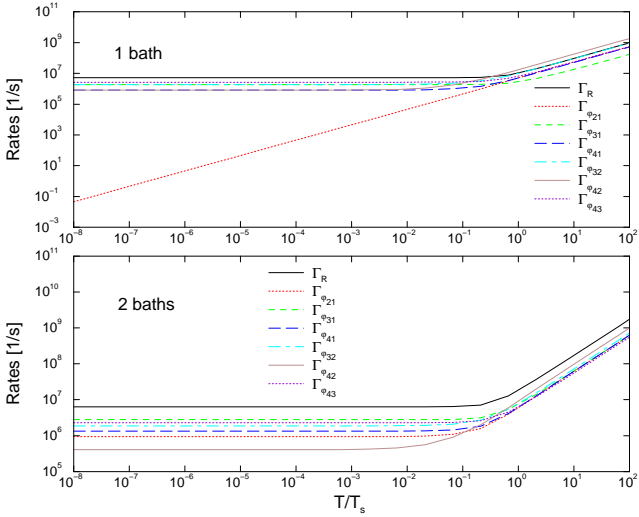}
\end{center}
\caption[Log-log plot of the temperature dependence of the sum of the
four relaxation rates and selected dephasing rates.]{Log-log plot of
the temperature dependence of the sum of the four relaxation rates and
selected dephasing rates. Qubit parameters $K$, $\epsilon$ and $\eta$
are all set to $E_s$ and $\alpha=10^{-3}$. The upper panel shows the
case of one common bath, the lower panel the case of two distinct
baths. At the characteristic temperature of approximately $0.1\cdot
T_s$ the rates increase very steeply.}
\label{fig_temp_dependence_1}
\end{figure}

Figure \ref{fig_temp_dependence_1} displays the dependence of typical
dephasing rates and the relaxation rate $\Gamma_R$ on temperature.
These decoherence rates are the inverse decoherence times.  The rates
are of the same magnitude for the case of one common bath and two
distinct baths.  As a notable exception, in the case of one common
bath the dephasing rates $\Gamma_{\varphi_{21}}=\Gamma_{\varphi_{12}}$
go to zero when the temperature  is decreased while all other rates
saturate for $T \rightarrow 0$. This phenomenon is explained later
on. If the temperature is increased from $T_s = (h/k_B) \cdot \nu_s = 4.8 \cdot
10^{-2}\textrm{ K}$ the increase of the dephasing and relaxation rates
follows a power law dependence. It is linear in temperature $T$ with a
slope given by the prefactors of the expression in the Redfield rates
that depends on temperature.  At temperature $T \approx 0.1 \cdot E_s$,
the rates show a sharp increase for both cases. This roll-off point is
set by the characteristic energy scale of the problem, which in turn is
set by the energy bias $\epsilon$, the transmission matrix element
$\eta$ and the coupling strength $K$. For the choice of parameters in
figure \ref{fig_temp_dependence_1} the characteristic energy scale
expressed in temperature is $T \approx 0.1 \cdot E_s$.

Note that there is also dephasing between the singlet and the triplet
states. When the system is prepared (by application of a suitable
interaction) in a coherent superposition of singlet and triplet states
the phase evolves coherently. Then two possible decoherence mechanisms
can destroy phase coherence. Firstly, ``flipless'' dephasing
processes, where $\langle E\rangle$ remains unchanged.  These flipless
dephasing processes are described by the terms for
$\omega_{lm},\omega_{nk} \rightarrow 0$ in the rates, eqs.\
(\ref{eq:rate_2bath}) and (\ref{eq:rate_1bath}).  Obviously these terms
vanish for $T \rightarrow 0$, as the low-frequency component of Ohmic
Gaussian noise is strictly thermal. Secondly, relaxation due to
emission of a boson to the bath is also accompagnied by a loss of
phase coherence. This process in  general has a {\em finite} rate at
$T=0$. This explains the T-dependence of the rates in the single-bath
case: $\ket{E1}$ alone is protected from the environment. As there are
incoherent transitions between the  triplet-eigenstates even at $T=0$,
the relative phase of a coherent oscillation betweeen $\ket{E1}$ and
any of those is randomized and  the decoherence rates
$\Gamma_{\varphi_{3/4, 1}}$ are finite even at $T=0$.  As a notable
exception, $\ket{E2}$, the lowest-energy state in the triplet
subspace, can only be flipped through absorption of energy, which
implies that the dephasing rate $\Gamma_{\varphi_{21}}$  also vanishes
at low temperature.  The described behaviour can be observed in figure
\ref{fig_temp_dependence_1}.

\begin{figure}[ht]
\begin{center}
\includegraphics[width=8.5cm]{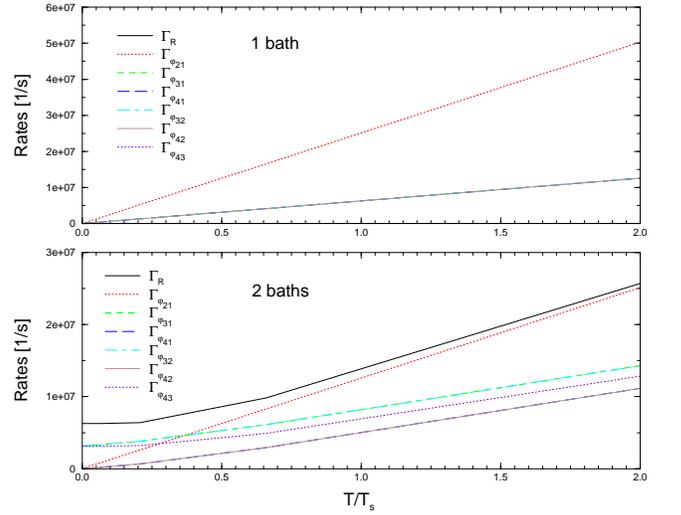}
\end{center}
\caption[Plot of the temperature dependence of the sum of the four
relaxation rates and selected dephasing rates.]{Plot of the
temperature dependence of the sum of the four relaxation rates and
selected dephasing rates. Qubit parameters $\epsilon$ and $\eta$ are
set to 0, $K$ is set to $E_s$, $\alpha=10^{-3}$ corresponding to the
choice of parameters used for the UXOR operation. The upper panel shows
the case of one common bath, the lower panel the case of two distinct
baths. In the case of one common bath the system will experience no
dissipative effects at $T=0$.}
\label{fig_temp_dependence_2}
\end{figure}

If the parameters $\epsilon$ and $\eta$ are tuned to zero, thus $K$
being the only non-vanishing parameter in the Hamiltonian, all
dephasing and relaxation rates will vanish for $T=0$ in the case of
one common bath.  This behaviour is depicted in figure
\ref{fig_temp_dependence_2}. It originates from the special symmetries
of the Hamiltonian in this case and the fact that for this particular
two qubit operation the system Hamiltonian and the coupling to the
bath are diagonal in the same basis.  This special case is of crucial
importance for the quantum gate operation as  described in
\ref{ch_gate_quality_factors} and affects the gate quality factors.


\subsection{Renormalization effects} \label{renormalization_effects}
\begin{figure}
\begin{center}
\includegraphics[width=8.6cm]{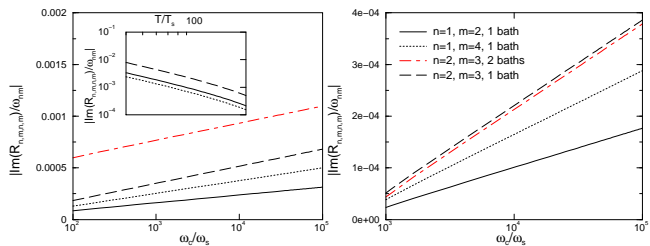}
\end{center}
\caption[Illustration of the size of the renormalization
effects.]{The left plot depicts the ratio of the
renormalization effects and the corresponding transition
frequencies. Parameters: $\alpha=10^{-3}$, $T=0$ and $\omega_c/\omega_s$
is varied between $10^{2}$ and $10^{5}$ for several frequencies
($\omega_{12}$, $\omega_{14}$ and $\omega_{23}$) for the case of two
baths and in the case of $\omega_{23}$ also for the case of one common
bath. The parameters for the right plot are $\alpha=10^{-3}$,
$T=2.1\cdot E_s$ and $\omega_c/\omega_s$ is varied between $10^{3}$ and
$10^{5}$. The inset of the left plot shows a log-log plot of the
temperature dependence of the renormalization effects. Here $\alpha=10^{-3}$ and
$\omega_c=10^{13}$. Note that for small temperatures the
renormalization effects do \textit{not} depend on temperature.  (This
is elucidated further in chapter \ref{renormalization_effects}.)  The plots
are scaled logarithmically to emphasize the logarithmic
divergence of the renormalization effects with $\omega_c$.}
\label{fig_renormalization}
\end{figure}

Next to causing decoherence, the interaction with the bath also
renormalizes the qubit frequencies. This is mostly due to the fast
bath modes and  can be understood analogous to the Franck-Condon
effect, the Lamb shift, or the adiabatic renormalization
\cite{Leggett}.  Renormalization of the oscillation frequencies
$\omega_{nm}$ is controlled by the imaginary part of the Redfield
tensor \cite{governale}
\begin{equation}
\omega_{nm} \rightarrow \tilde \omega_{nm} :=
\omega_{nm}-\mbox{Im}R_{nmnm}\textrm{.}
\end{equation}
Note that $\mbox{Im}R_{nmnm}=-\mbox{Im}R_{mnmn}$ due to the fact that
the correlators in the Golden Rule expressions have the same parity.
The imaginary part of the Redfield tensor is given by
\begin{eqnarray}
\mbox{Im} \ \Gamma^{(+)}_{\ell m nk} & = & C^{\rm 1b,2b}_{\ell m n k} \frac{1}{\pi
\hbar} {\cal P}\int_{0}^\infty \limits d\omega \ J(\omega)
\left(\frac{1}{\omega^2-\omega_{nk}^2}\right) {}
\nonumber \\ & &{} \times \left[ \mbox{coth}(\beta \hbar
\omega/2)\omega_{nk}-\omega\right]
\end{eqnarray}
and
\begin{eqnarray}
\mbox{Im} \ \Gamma^{(-)}_{\ell m nk} & = & C^{\rm 1b,2b}_{\ell m n k} \frac{1}{\pi
\hbar} {\cal P} \int_{0}^\infty \limits d\omega \ J(\omega)
\left(\frac{1}{\omega^2-\omega_{lm}^2}\right) {}
\nonumber \\ & &{} \times \left[ \mbox{coth}(\beta \hbar
\omega/2)\omega_{lm}+\omega\right] \textrm{,}
\end{eqnarray}
where ${\cal P}$ denotes the principal value and $C^{\rm 1b,2b}_{\ell m n k}$ are pre-factors
defined, in the case of two distinct baths, according to $C^{\rm 2b}_{\ell m n k} =
\frac{1}{4}\left[ \sigma_{z,\ell
m}^{(1)}\sigma_{z,nk}^{(1)}+\sigma_{z,\ell m}^{(2)}
\sigma_{z,nk}^{(2)}\right]$ and in the case of one common bath $C^{\rm 1b}_{\ell m n k}
= \frac{1}{4} \Lambda$. Here, for simplicity, we assumed $\alpha_1=
\alpha_2=\alpha$ and thus $J_1(\omega)=J_2(\omega)=J(\omega)$.
Evaluation of the integral leads to the
following expression for $\Gamma^{(+)}_{\ell mnk}$
\begin{eqnarray}
\mbox{Im} \ \Gamma^{(+)}_{\ell mnk} & = & C^{\rm 1b,2b}_{\ell m n k} \frac{\alpha
	\omega_c^2 \omega_{nk}}{2\pi (\omega_c^2+ \omega_{nk}^2)}
	\Bigg [\psi(1+c_2) + \psi(c_2) {} \nonumber \\ & &{} -
	2\mbox{Re}[\psi(ic_1)] - \pi \frac{\omega_c}{\omega_{nk}}
	\Bigg ]\textrm{,}\label{impart_gammap}
\end{eqnarray}
with $c_1:=(\beta \hbar \omega_{nk})/(2\pi)$, $c_2:=(\beta\hbar
\omega_c)/(2\pi)$.  In the case of $\Gamma^{(-)}_{\ell mnk}$ the
expression is
\begin{eqnarray}
\mbox{Im} \ \Gamma^{(-)}_{\ell mnk} & = & C^{\rm 1b,2b}_{\ell m n k} \frac{\alpha
	\omega_c^2 \omega_{\ell m}}{2\pi (\omega_c^2+ \omega_{\ell
	m}^2)} \Bigg [\psi(1+c_2) + \psi(c_2) {} \nonumber \\ & &{}
	- 2\mbox{Re}[\psi(ic_1)] + \pi \frac{\omega_c}{\omega_{\ell
	m}} \Bigg ]\textrm{,}\label{int_final_Gammam}
\end{eqnarray}
with $c_1:=(\omega_{\ell m}\beta\hbar)/(2\pi)$.  The terms in
(\ref{impart_gammap}) and (\ref{int_final_Gammam}) which are linear in
$\omega_c$ give no net contribution to the imaginary part of the
Redfield tensor \cite{governale}.  To illustrate the size of the
renormalization effects, the ratio of the renormalization effects to
the frequencies which are renormalized is depicted in figure
\ref{fig_renormalization}.

If $c_1$ and $c_2$ are large and the digamma functions can be
approximated by a logarithm the resulting expression for the
renormalization effects will be independent of temperature. The
temperature dependence of (\ref{impart_gammap}) and
(\ref{int_final_Gammam}) at higher temperatures, where $c_1$ and $c_2$
are small and the renormalization effects are very weak is shown in
figure \ref{fig_renormalization}.  The rates (\ref{impart_gammap}) and
(\ref{int_final_Gammam}) diverge logarithmically with $\omega_c$ in
analogy to the well-known ultraviolet-divergence of the spin Boson
model \cite{Leggett}.  When comparing the upper left ($T=0$) and upper
right ($T=2.1\cdot E_s$) panel, one recognizes that for the first case
one common bath gives somewhat smaller renormalization effects than
two distinct baths, while in the second case for $T=2.1\cdot E_s$ the
renormalization effects deviate only slightly (see the behaviour for
$\omega_{23}$) and the renormalization effects are smaller for the
case of two distinct baths.  The effects of renormalization are always
very small ($\vert \mbox{Im}(R_{n,m,n,m})/\omega_{nm}\vert$ below 1\%
for our choice of parameters) and are therefore
neglected in our calculations. However, having calculated
(\ref{impart_gammap}) and (\ref{int_final_Gammam}), these are easily
incorporated in our calculation. The case of large renormalization
effects is discussed in Ref.\ \onlinecite{kehrein}.

We only plotted the size of the renormalization effects for
$\omega_{12}$, $\omega_{14}$ and $\omega_{34}$, because in general all
values of $\omega_{nk}$ are of the same magnitude and give similar
plots. The
size of the renormalization effects diverges linearly with $\alpha$,
the dimensionless parameter which describes the strength of the
dissipative effects.

For flux qubits, the cutoff frequency $\omega_c$ is given by the
circuit properties. For a typical first order low-pass LR filter in a
qubit circuit \cite{caspar}, one can insert $R=50$~$\Omega$ (typical
impedance of coaxial cables) and $L \approx 1$ nH (depends on the
length of the circuit lines) into $\omega_{LR}=R/L$ and gets that
$\omega_{LR} \approx 5 \cdot 10^{10}$ Hz. $\omega_{LR}$ is the largest
frequency in the problem (see again \cite{caspar}, chapter 4.5) and
$\omega_c \gg \omega_{LR}$ should hold. Then $\omega_c \approx
10^{13}$ Hz (=$10^4\cdot E_s$) as cutoff frequency is a reasonable
assumption.

\section{Gate Quality Factors} \label{ch_gate_quality_factors}
In the last section \ref{ch_bloch_redfield_formalism}, we evaluated the
dephasing and relaxation rates of the two-qubit system that is
affected by a dissipative environment. Furthermore, we visualized the
dynamics of the two-qubit system. This does not yet allow a full
assessment of the performance as a quantum logic element. These should
perform unitary gate operations and based on the rates alone, one can not judge
how well quantum gate operations might be performed with the two qubit
system. Therefore, to get a quantitative measure of how our setup 
behaves when performing a quantum logic gate operation, one can
evaluate gate quality factors \cite{GateQualityFactors}. The
performance of a two-qubit gate is characterized by four quantities:
the fidelity, purity, quantum degree and entanglement capability. The
fidelity is defined as
\begin{equation}
\mathcal{F} = \frac{1}{16} \sum_{j=1}^{16}
\braket{\Psi_{in}^j|U_{G}^+\rho^j_{G}U_{G}|
\Psi^j_{in}}\textrm{,}
\end{equation}
where the density matrix obtained from attempting a quantum gate operation in
a hostile environment is $\rho^j_{G}=\rho(t_{G})$, which is
evaluated for all initial conditions
$\rho(0)=\ket{\Psi_{in}^j}\bra{\Psi_{in}^j}$. The fidelity is a
measure of how well a quantum logic operation was performed. Without
dissipation the reduced density matrix $\rho^j_{G}$ after performing
the quantum gate operation, applying $U_{G}$ and the inverse
$U_{G}^+$ would equal $\rho(0)$. Therefore the fidelity for the
ideal quantum gate operation should be 1.

The second quantifier is the purity
\begin{equation}
\mathcal{P} = \frac{1}{16} \sum_{j=1}^{16}
\mbox{tr}\left((\rho^j_{G})^2\right)\textrm{,}
\end{equation}
which should be 1 without dissipation and $1/4$ in a fully mixed
state.  The purity characterizes the effects of decoherence.

The third quantifier, the quantum degree, is defined as the maximum
overlap of the resulting density matrix after the quantum gate operation with
the maximally entangled states, the Bell-states
\begin{equation}
\mathcal{Q} = \max_{j,k}
\braket{\Psi_{me}^k|\rho_{G}^j|\Psi_{me}^k}\textrm{,}
\end{equation}
where the Bell-states $\Psi_{me}^k$ are defined according to
\begin{eqnarray} \label{bell_states}
\ket{\Psi_{me}^{00}} & = & \frac{\ket{\down \down} +
\ket{\up\up}}{\sqrt{2}}\textrm{, } \ket{\Psi_{me}^{01}}  =
\frac{\ket{\down \up} + \ket{\up \down}}{\sqrt{2}}\textrm{,} \\
\ket{\Psi_{me}^{10}} & = & \frac{\ket{\down \down} - \ket{\up
\up}}{\sqrt{2}}\textrm{, } \ket{\Psi_{me}^{11}}  =  \frac{\ket{\down
\up} - \ket{\up \down}}{\sqrt{2}}\textrm{.}
\end{eqnarray}
For an ideal entangling operation, e.g. the UXOR gate, the quantum
degree should be 1. The quantum degree characterizes nonlocality. It has been
shown \cite{nonlocality} that all density operators that have an overlap with
a maximally entangled state that is larger than the value $0.78$ \cite{thorwart}
violate the Clauser-Horne-Shimony-Holt inequality and are thus
non-local.

The fourth quantifier, the  entanglement capability $\mathcal{C}$, is
the smallest eigenvalue of the partially transposed density matrix for
all possible unentangled input states $\ket{\Psi_{in}^j}$. (see below).
 It has been shown \cite{peres} to
be negative for an entangled state. This quantifier should be
-0.5, e.g. for the ideal UXOR, thus characterizing a maximally entangled
final state. Two of the gate quality factors, namely the fidelity and purity
might also be calculated for single qubit gates
\cite{marlies}. However, entanglement can only be observed in a system
of at least two qubits. Therefore, the quantum degree and entanglement
capability cannot be evaluated for single qubit gates.

To form
all possible initial density matrices, needed to calculate the gate
quality factors, we use the 16 unentangled product states
$\ket{\Psi_{in}^j}$, $j=1,\dots ,16$ defined \cite{thorwart} according
to $\ket{\Psi_a}_1\ket{\Psi_b}_2$, ($a,b=1,\dots,4$),  with
$\ket{\Psi_1}=\ket{\down}$, $\ket{\Psi_2}=\ket{\up}$,
$\ket{\Psi_3}=(1/\sqrt{2})(\ket{\down}+ \ket{\up})$, and
$\ket{\Psi_4}=(1/\sqrt{2}) (\ket{\down}+i\ket{\up})$. They form one possible
basis set for the superoperator $\nu_{G}$ with
$\rho(t_{G})=\nu_{G} \rho(0)$   \cite{GateQualityFactors,
thorwart}. The states are choosen to be unentangled for being compatible
with the definition of $\mathcal{C}$. 

\subsection{Implementation of two-qubit operations}

\subsubsection{Controlled phase-shift gate}
To perform the UXOR operation, it is necessary to be able to
apply the controlled phase-shift operation together with arbitrary
single-qubit gates. In the computational basis ($\ket{00}$, $\ket{01}$,
$\ket{10}$, $\ket{11}$), the controlled phase-shift operation is given by
\begin{eqnarray}
U_{CZ}(\varphi) & = &  \left( \begin{array}{cccc} 
1 & 0 & 0 & 0 \\ 
0 & 1 & 0 & 0 \\
0 & 0 & 1 & 0 \\
0 & 0 & 0 & e^{i\varphi} \\
\end{array} \right)\textrm{,}
\end{eqnarray}
and for $\varphi=\pi$, up to a global phase factor,
\begin{equation} \label{ucz}
U_{CZ} = \exp\left(i \frac{\pi}{4} \sigma_z^{(1)}\right)
	\exp\left(i \frac{\pi}{4}\sigma_z^{(2)}\right) 
	\exp\left(i \frac{\pi}{4}\sigma_z^{(1)} \sigma_z^{(2)}\right)\textrm{.}
\end{equation}
Note that in (\ref{ucz}) only $\sigma_z$-operations, which commute with the coupling to
the bath, are needed. The controlled phase-shift operation together with two Hadamard
gates and a single-qubit phase-shift operation then gives the UXOR gate.

\subsubsection{UXOR gate}
Due to the fact that the set consisting of the UXOR (or CNOT) gate and the
one-qubit rotations, is complete for quantum computation
\cite{barenco}, the UXOR gate is a highly important two-qubit gate
operation. Therefore we further investigate the behaviour of the four
gate quality factors in this case. The UXOR operation switches the second bit,
depending on the value of the
first bit of a two bit system. In the computational basis,
this operation has the following matrix form
\begin{eqnarray} \label{xor_matrix_compbasis}
U_{XOR} & = &   \left( \begin{array}{cccc} 1 & 0 & 0 & 0 \\ 0 & 1 & 0
& 0 \\ 0 & 0 & 0 & 1 \\ 0 & 0 & 1 & 0 \\
\end{array} \right)\textrm{.}
\end{eqnarray}
Up to a phase factor, the two-qubit UXOR (or CNOT) operation can be realized by
a sequence of five single-qubit and one two-qubit quantum
logic operation. Each of these six operations corresponds to an
appropriate Hamiltonian undergoing free unitary time evolution
$\exp(-(i/\hbar) \mathbf{H}_{op}t)$. The single-qubit operations are
handled with Bloch-Redfield formalism, like the two-qubit
operations. We assume dc pulses (instantaneous on and off switching of
the Hamiltonian with zero rise time of the signal) or rectangular
pulses
\begin{eqnarray} \label{uxor}
U_{XOR} & = &	\exp\left(-i\frac{\pi}{2}\left(\frac{\sigma_x^{(2)}
		+\sigma_z^{(2)}}{\sqrt{2}}\right)\right) {} \nonumber \\
		& &{} \times  U_{CZ}(\pi) {} \nonumber \\
		& &{} \times \exp \left( i \frac{\pi}{2} \sigma_z^{(1)} \right) {} \nonumber\\
		& &{} \times \exp\left(-i\frac{\pi}{2}\left(\frac{\sigma_x^{(2)}
		+\sigma_z^{(2)}}{\sqrt{2}}\right)\right)\textrm{,}
\end{eqnarray}
where $U_{CZ}(\pi)$ is given by (\ref{ucz}).
This generic implementation has been chosen in order to demonstrate the comparison to other coupling schemes \cite{thorwart} as well as for computational
convenience, it is not necessarily the optimum scheme for application under
cryogenic conditions, where slow rise-time AC pulses are preferred.
Table \ref{xorparam} shows the parameters we inserted into the one-
and two-qubit Hamiltonian to receive the UXOR operation.
\begin{table}
\begin{center}
\begin{tabular}{|c|c|c|c|}
\hline No. & operation & parameters [$E_s$] & time [s] \\ \hline
\hline 1   &
$\exp\left(-i\frac{\pi}{2}\left(\frac{\sigma_x^2+\sigma_z^2}{\sqrt{2}}\right)\right)$
& $\epsilon_2=-\xi$, $\Delta_2=-\xi $ &
$\tau_1=\frac{\sqrt{2}}{2\xi}$\\ \hline 2   & $\exp\left(i
\frac{\pi}{2} \sigma_z^1\right)$ & $\epsilon_1=\xi$ &
$\tau_2=\frac{1}{2\xi}$\\ \hline 3   & $\exp\left(i \frac{\pi}{4}
\sigma_z^1 \sigma_z^2\right)$ & $K=\xi$ & $\tau_3=\frac{1}{4\xi}$\\
\hline 4   & $\exp\left(i \frac{\pi}{4} \sigma_z^2\right)$ &
$\epsilon_2=\xi$ & $\tau_4=\frac{1}{4\xi}$\\ \hline 5   & $\exp\left(i
\frac{\pi}{4} \sigma_z^1\right)$ & $\epsilon_1=\xi$ &
$\tau_5=\frac{1}{4\xi}$\\ \hline 6   &
$\exp\left(-i\frac{\pi}{2}\left(\frac{\sigma_x^2+\sigma_z^2}{\sqrt{2}}\right)\right)$
& $\epsilon_2=-\xi$, $\Delta_2=-\xi$ &
$\tau_6=\frac{\sqrt{2}}{2\xi}$\\ \hline
\end{tabular}
\end{center}
\caption{Parameters of the Hamiltonians which are needed to perform the UXOR gate operation,
only the non-zero parameters are listed; $\xi=E_s$ in our case.}
\label{xorparam}
\end{table}
In our case we assumed $\xi=E_s$. However, there is no restriction in
the use of other values for $\xi$. For a typical energy
scale of 1 GHz, the resulting times from table
\ref{xorparam} are in the nanosecond range.
\begin{figure}[t]
\begin{center}
\includegraphics[width=8.5cm]{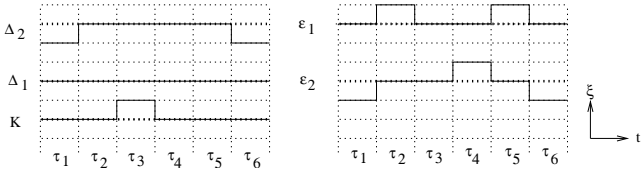}
\end{center}
\caption[Pulse sequence needed to perform the quantum UXOR
operation.]{Pulse sequence needed to perform the quantum UXOR
operation. Here the elements of the unperturbed single- and two-qubit
Hamiltonian needed to perform a certain operation undergoing free
unitary time evolution are shown. The dotted horizontal
lines denote $\xi=0$, and the
horizontal lines are spaced by $\vert \xi \vert=E_s$. The durations of
each pulse are not equal in general $\tau_j \neq \tau_i$,
$i,j=1,\ldots,6$ (cf. table \ref{xorparam})} \label{puls_shape_XOR}
\end{figure}

To better visualize the pulse sequence needed to perform the
quantum UXOR operation, which was already given in table \ref{xorparam},
figure \ref{puls_shape_XOR} depicts the values of the elements of the
Hamiltonians. Interestingly enough we find that for the only two-qubit
operation included in the UXOR operation $\epsilon$ and $\eta$ are
zero. Thus $K$ is the only non-zero parameter and $\mathbf{H}_{2qb}$
assumes diagonal form. For flux qubits, implementing the pulse sequence
figure \ref{puls_shape_XOR} involves negative and positive values 
tuning the magnetic frustration through the qubit
loop below or above $f=1/2$. Note that e.g. for realistic
models of inductively coupled flux qubits, it is very difficult to turn
on the interaction Hamiltonian between the two qubits without the
individual $\sigma_z$ terms in the Hamiltonian. However, for the pulse
sequence given in (\ref{uxor}), we might simply perform the third,
fourth and fifth operation of (\ref{uxor}) at once using only the
Hamiltonian with both the individual $\sigma_z$ terms and the
inter-qubit coupling.

To obtain the final reduced density matrix after performing the six
unitary operations (\ref{uxor}), we iteratively determine the density
matrix after each operation with Bloch-Redfield theory and insert the
attained resulting density matrix as initial density matrix into the
next operation. This procedure is repeated for all possible
unentangled initial states given in the last section. We inserted no
additional time intervals between the operations. This is usually
needed, if one applies Bloch-Redfield formalism, because
it is known to violate complete positivity on short
timescales. However, we circumvent this problem in our calculations by
dropping the memory after each operation, when we iteratively
calculate the reduced density matrix. This procedure may lead to 
small inaccuracies as compared to using QUAPI \cite{thorwart}, which
however should not affect our main conclusions.

\subsection{Temperature dependence}

\subsubsection{Controlled phase-shift gate}
\begin{figure}[t]
\begin{center}
\includegraphics[width=8.5cm]{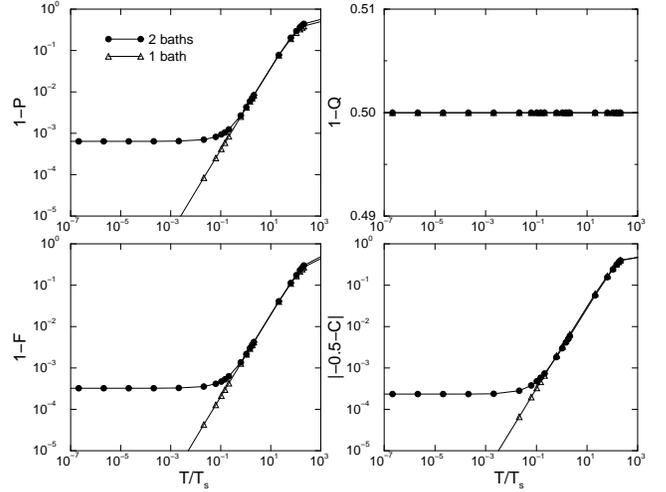}
\end{center}
\caption[Log-log plot of the temperature dependence of the deviations
of the four gate quantifiers from their ideal values.]{Log-log plot of
the temperature dependence of the deviations of the four gate
quantifiers from their ideal values after performing the controlled phase-shift
(CPHASE) gate operation. Here the temperature is varied
from $\approx 0 $  to  $2\cdot E_s$. In all cases
$\alpha=\alpha_1=\alpha_2=10^{-3}$.}
\label{gqfTdepcz}
\end{figure}

We have analyzed the gate quality factors in the cases of a common and
of two distinct baths respectively. 
In figure \ref{gqfTdepcz} the temperature dependence of the deviations of
the four gate quality factors
from their ideal values are depicted as a log-log plot.
At temperatures below $T=2.5 \cdot10^{-2}$~K $ \approx 0.5\cdot T_s$, the purity
and fidelity are clearly higher for the case of one common bath, but if temperature is
increased above this characteristic threshold, fidelity and purity are
slightly higher for the case of two baths.

In the case of one common bath the fidelity, purity and entanglement
capability are approaching their ideal value 1, when temperature goes to
zero.  This is related to the fact that in the case of one common bath all
relaxation and dephasing rates vanish during the two-qubit step of the
controlled phase-shift gate due to the special symmetries of the
Hamiltonian, when temperature goes to zero as depicted in figure
\ref{fig_temp_dependence_2}. 

The controlled phase-shift operation creates
entanglement. The creation of entanglement is impeded by decoherence effects
that vanish when temperature approaches zero. Therefore, the entanglement
capability exhibits the same behaviour as the fidelity and purity.
For zero dissipation ($\alpha=0$) the quantum degree has the value 0.5 but
the entanglement capability is -0.5 thus characterizing a maximum entangled
state. The reason is that the Bell-states, which are generated by the controlled
phase-shift gate from the input states, result in a basis that is different
from the used basis, but can be transformed using only local transformations.

Furthermore, for finite dissipation, figure \ref{fig_temp_dependence_2} shows
that also for the case of two distinct baths, there are only three non-vanishing rates for $T
\rightarrow 0$. The system, being prepared in one of the 16 initial
states, might relax into one of the eigenstates that is an entangled
state.

We observe the saturation of the deviation for the case of
two baths and can directly recognize the effects of the symmetries of
the controlled phase-shift operation. For given
$\alpha$, the fidelity and purity cannot be increased anymore by
lowering the temperature in the case of two distinct
baths. Interestingly enough, we find that for two qubits coupling
to one common bath, the situation is different for temperatures
below $0.5 \cdot T_s$. Above a
temperature of $T_s = 4.8 \cdot 10^{-2}$ K, the decrease of the gate quality factors
shows a linear dependence on temperature for both cases of one
common or two distinct heat baths before it again saturates at
about $10^{2}$~K $\approx 10^4 \cdot T_s$.  Finite
decoherence effects in the fidelity, purity and entanglement
capability at $T=0$ for the case of two distinct baths are resulting
from the coupling of the system to the environment of harmonic oscillators,
which (at $T=0$) are all in their ground states and can be excited through
spontaneous emission. But for the case of one common bath,
the deviation from the ideal fidelity goes to zero, when temperature
goes to zero. This is due to the special symmetries ($K$ is the only
\textit{non}vanishing parameter in the two-qubit operation) of the
Hamiltonian, which rules out spontaneous emission. 
These symmetries are also reflected in the temperature
dependence of the rates, figure \ref{fig_temp_dependence_2}. There,
for one common bath, all rates vanish for $T\rightarrow 0$.
Note that these rates only describe the two-qubit part of the operation. However,
the single-qubit part behaves similarly because the terms in the single-qubit
Hamiltonian are als $\propto \sigma_z$.

\subsubsection{UXOR gate}
\begin{figure}[t]
\begin{center}
\includegraphics[width=8.5cm]{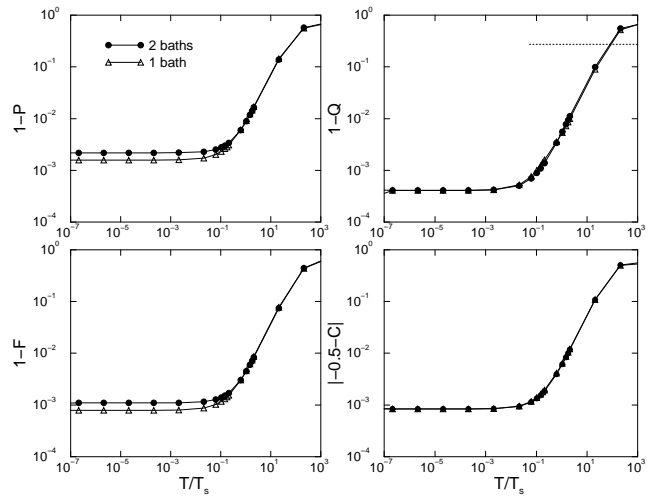}
\end{center}
\caption[Log-log plot of the temperature dependence of the deviations
of the four gate quantifiers from their ideal values.]{Log-log plot of
the temperature dependence of the deviations of the four gate
quantifiers from their ideal values after performing the UXOR gate
operation. Here the temperature is varied
from $\approx 0 $  to  $2\cdot E_s$. In all cases
$\alpha=\alpha_1=\alpha_2=10^{-3}$. The dotted line indicates the
upper bound set by the Clauser-Horne-Shimony-Holt inequality.}
\label{gqfTdep}
\end{figure}

Different to the last section, we now add two single-qubit operations (Hadamard
gates) to the controlled phase-shift operation that do \textit{not} commute with
the coupling to the bath.
In figure \ref{gqfTdep}, the deviations of the gate quality factors from their
ideal values are depicted as a log-log plot.
Again, at temperatures below $T=2.5
\cdot10^{-2}$~K$ \approx 0.5\cdot E_s$, the purity and fidelity are
higher for the case of one common bath, but if temperature is
increased above this characteristic threshold, fidelity and purity are
higher for the case of two baths. Note that we have chosen a rather large
$\alpha$, this value can substantially be improved by means of engineering 
\cite{caspar}.

The fidelity and purity are clearly higher for the case of one
common bath, when temperature is decreased below $0.5 \cdot T_s$. This is related
to the fact that in the case of one common bath all relaxation and
dephasing rates vanish during the two-qubit-step of the UXOR, 
due to the special symmetries of the
Hamiltonian, when temperature goes to zero as discussed in the
last paragraph. However, the quantum degree and the
entanglement capability tend towards the same value for both the
case of one common and two distinct baths. This is due to the fact
that both quantum degree and entanglement capability are, different
than fidelity and purity, not defined as mean values but rather
characterize the ``best'' possible case of all given input
states. This results in the same value for both cases.

In the recent work by Thorwart and H\"anggi \cite{thorwart}, the UXOR 
gate was investigated for a $\sigma_y^{(i)}\otimes\sigma_y^{(j)}$ coupling scheme 
and one common bath. They find a pronounced degradation of the
gate performance, in particular, the gate quality factors only
weakly depend on temperature. 
They set the strength of the dissipative effects to
$\alpha=10^{-4}$. Their choice of parameters was $\epsilon \approx 10
\cdot E_s$, $\Delta \approx 1 \cdot E_s$ and $K \approx 0.5 \cdot E_s$
which is on the same order of magnitude as the values given in table
\ref{xorparam}.  As can be seen in 
figure \ref{gqfTdep}, we also observe only a weak decrease
of the gate quality factors for both the case of one common bath and
two distinct baths in the
same temperature range discussed by Thorwart and H\"anggi, both for
$\alpha=10^{-3}$ and $\alpha=10^{-4}$ and overall substantially better values.
This is due to the fact that for $\sigma_y^{(i)}\otimes\sigma_y^{(j)}$
coupling, the Hamiltonian does {\em not} commute with the coupling to the
bath during the two-qubit steps of the UXOR pulse sequence. 

We observe the saturation of the deviation for both the case of
two baths and one common bath. For given
$\alpha$, the fidelity and purity can not be increased anymore by
lowering the temperature, different from the behaviour for the
controlled phase-shift gate that was discussed in the last
section. This is due to the application of the Hadamard gate whose
Hamiltonian does not commute with the coupling to the bath. Above a
temperature of $10^{-2}$ K, the decrease of the gate quality factors
shows a linear dependence on temperature for both cases.  Here,
different from the controlled phase-shift gate, we observe finite
decoherence effects in all four gate quantifiers also at $T=0$,
both for the case of one common or two distinct heat baths. These
decoherence effects are resulting from the coupling of the
system to the environment of harmonic oscillators, which (at $T=0$)
are all in their ground states and can be excited through spontaneous
emission as already described above.

The dotted line in figure \ref{gqfTdep} shows that the temperature has
to be less than about $T=21 \cdot E_s=1$~K in order to obtain values of the quantum
degree being larger than $\mathcal{Q} \approx 0.78$. Only then, the
Clauser-Horne-Shimony-Holt inequality is violated and non-local
correlations between the qubits occur as described in \cite{thorwart}.

\subsection{Dependence on the dissipation strength} \label{ch_gqf_dissipation_strength_dependence}
The deviations from the ideal values of the gate quantifiers possess a
linear dependence on $\alpha$ as expected. 
Generally (if no special
symmetries of the Hamiltonian are present) there are always finite
decoherence effects also at $T=0$.
Therefore, we can not improve the gate quality
factors below a certain
saturation value, when lowering the temperature \cite{thorwart},
as was also discussed in the last section.
By better isolating
the system from the environment and by carefully engineering
the environment \cite{caspar} one can decrease the strength of the
dissipative effects characterized by $\alpha$.
In order to obtain
the desired value of 0.999 99 for $\mathcal{F}$, $\mathcal{P}$ and
$\mathcal{Q}$ \cite{thorwart}, $\alpha$ needs to be below
$\alpha=10^{-6}$ at $T=0.21\cdot E_s$.

\subsection{Time resolved UXOR operation}
To investigate the anatomy of the UXOR quantum logic operation, we
calculated the occupation probabilities of the singlet/triplet states
after each of the six operations, of which the UXOR consists. This time
resolved picture of the dynamics of the two-qubit system, when
performing a gate operation, gives insight into details of our
implementation of the UXOR operation and the dissipative effects
occuring during the operation. Thus we are able to completely
characterize the physical process, which maps the input density matrix
$\rho_0$ to $\rho_{out}$ in an open quantum system
\cite{GateQualityFactors}.  When the system is prepared in the state
$\ket{\down \down}=\ket{00}$ the UXOR operation
(\ref{xor_matrix_compbasis}) does not alter the initial state and
after performing the UXOR operation the final state should equal the
initial state $\ket{\down \down}=\ket{00}$.
\begin{figure}[t]
\begin{center}
\includegraphics[width=8.5cm]{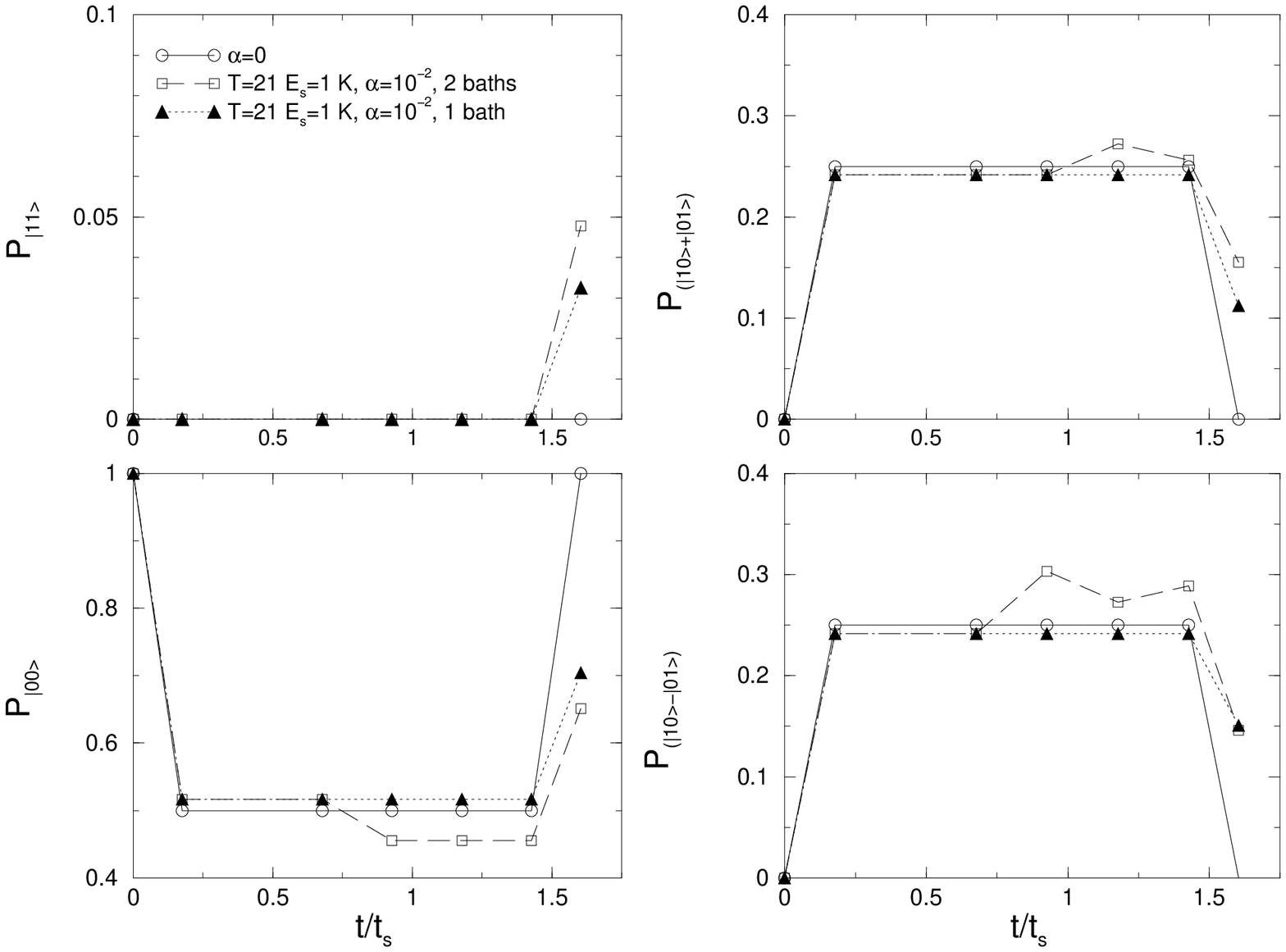}
\end{center}
\caption[Time resolved UXOR operation. The system is initially prepared
in the state $\ket{0 0}$.]{Time resolved UXOR operation. The system is initially
prepared in the state $\ket{0 0}$.\ Occupation
probabilities of the singlet/triplet states are shown after completion
of a time step $\tau_i$ ($i=1,\ldots,6$).  For
$\alpha=\alpha_1=\alpha_2=10^{-2}$ and $T=21\cdot E_s=1$~K clearly
deviations from the ideal case can be observed. Qubit parameters are
set according to table \ref{xorparam}.}
\label{time_resolved_xor_downdown}
\end{figure}
This can clearly be observed in figure
\ref{time_resolved_xor_downdown}.  During the UXOR operation occupation
probabilities of the four states change according to the individual
operations given in (\ref{uxor}).  At $T=21 \cdot E_s$, the case of 2
baths differs significantly from the case of one common bath. After
the third operation (the two-qubit operation; only there the
distinction between one common or two distinct baths makes sense)
occupation probabilities are different for both environments resulting
in a less ideal result for the case of two baths.

In figure \ref{time_resolved_xor_downdown}, the resulting state after
performing the UXOR operation always deviates more from the ideal value
(for $\alpha=0$, i.e.\ no dissipation) for the case of two distinct
baths, if all other parameters are fixed and set to the same values for
the both cases. The state $P_{\ket{00}}$ is less close to the ideal
occupation probability 1 and the other singlet/triplet states are also
less close to their ideal value for the case of two distinct
baths. The case of two distinct baths also shows bigger deviations
from the ideal case ($\alpha=0$) \textit{during} the UXOR operation
(see figure \ref{time_resolved_xor_downdown}). But, if the system is
initially prepared in the state $\ket{\up\up}=\ket{11}$ the case of
two distinct baths shows bigger deviations from the ideal case during
the UXOR operation, while the resulting state is closer to the ideal
case for two distinct baths compared to one common bath.

\begin{figure}[t]
\begin{center}
\includegraphics[width=8.5cm]{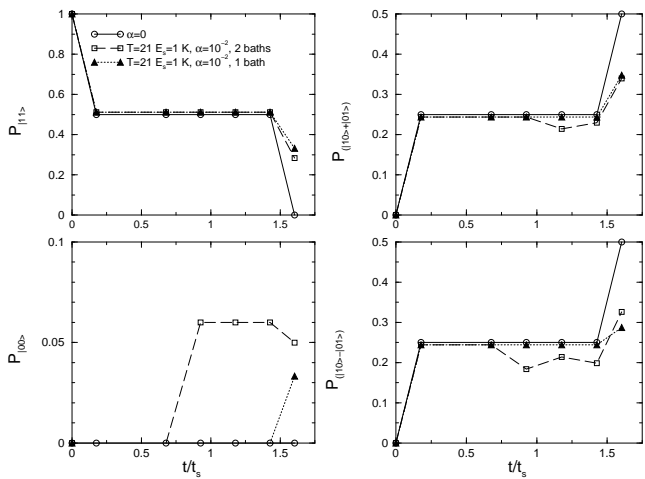}
\end{center}
\caption[Time resolved UXOR operation. The system is initially prepared
in the state $\ket{11}$.]{Time resolved UXOR operation. The system is initially
prepared in the state $\ket{11}$.\ Occupation
probabilities of the singlet/triplet states are shown after completion
of a time step $\tau_i$ ($i=1,\ldots,6$).  For
$\alpha=\alpha_1=\alpha_2=10^{-2}$ and $T=21 \cdot E_s=1$~K deviations from the
ideal case can be observed. Qubit parameters are set according to
table \ref{xorparam}.}
\label{time_resolved_xor_upup}
\end{figure}
\begin{figure}[t]
\begin{center}
\includegraphics[width=8.5cm]{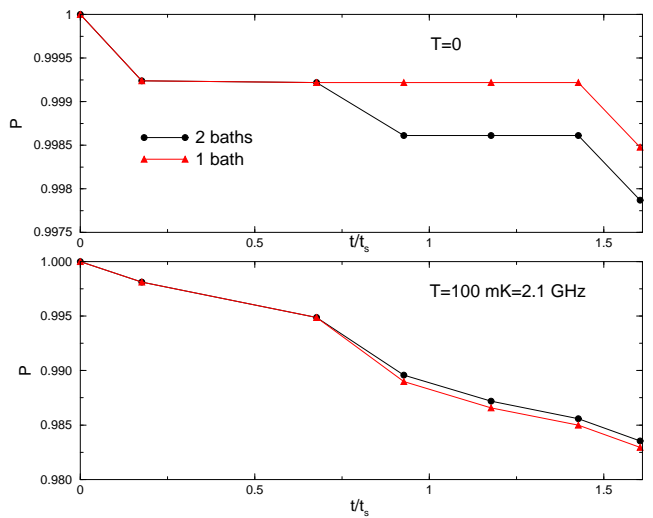}
\end{center}
\caption[Time resolved Purity during the quantum UXOR operation.]{Time
resolved Purity for the UXOR operation. The value of the purity after
each time step $\tau_i$ ($i=1,\ldots,6$) is shown. Here
$\alpha=\alpha_1=\alpha_2=10^{-2}$ and $T=2.1 \cdot E_s =100$~mK (lower panel)
or $T=0$ (upper panel). Qubit parameters are set according to table \ref{xorparam}.}
\label{time_resolved_P}
\end{figure}

In figures \ref{time_resolved_xor_downdown} and
\ref{time_resolved_xor_upup} it looks like there would be no
decoherence effects (or at least much weaker decoherence effects)
after performing the (first two) single-qubit operations. However, not
all input states are affected by the decoherence effects the same way.
And when we regard all possible input states, there are finite
decoherence effects. This can be explained with figure
\ref{time_resolved_P}.  Figure \ref{time_resolved_P} depicts the time
resolved purity when performing the UXOR operation. We clearly observe
that there are finite decoherence effects for the first single-qubit
operations in (\ref{uxor}) as well. The difference between the single-qubit and
two-qubit operations is the steeper decrease of the purity due to
stronger decoherence in the case of the two-qubit operation. The upper
panel in figure \ref{time_resolved_P} depicts the behaviour of the
purity for $T\rightarrow 0$. Decoherence due to the $\sigma_z$ terms
in the Hamiltonian will vanish for $T\rightarrow 0$ in the case of
one common bath.

\section{Conclusion}
We presented a full analysis of the dynamics and decoherence
properties of two solid state qubits coupled to each other via a
generic type of Ising coupling and coupled, moreover, either to a
common bath, or two independent baths.

We calculated the dynamics of the system and evaluated decoherence
times. From the temperature dependence of the decoherence rates
(figure \ref{fig_temp_dependence_1}), we conclude that both types of
environments show a similar behaviour; however, in the case of one
common bath, two of the decoherence rates are zero, and the remaining
ones are slightly larger than in the case of two distinct baths. This
temperature dependence is also reflected in the characteristics of the
gate quality factors from quantum information theory, which
are introduced as robust measures of the quality of a quantum logic
operation. We illustrate that the gate quality factors depend linearly
on $\alpha$, as expected.  The time resolved UXOR operation (figures
\ref{time_resolved_xor_downdown} and \ref{time_resolved_xor_upup})
again illustrates the difference between one common and two distinct
baths, and moreover we observe that single-qubit decoherence effects
$\propto \sigma_z$ during the UXOR operation are weak. The time scales of the dynamics of
the coupled two qubit system are comparable to the time scales, which
were already observed in experiments and discussed in the literature
\cite{caspar}.

The question, whether one common bath or two distinct baths are less
destructive regarding quantum coherence can not be clearly
answered. For low enough temperatures, coupling to one common
bath yields better results. However, when the temperature is increased
two distinct baths do better; in both temperature regimes, though, the
gate quantifiers are only slightly different for both cases.

Compared to the work of Thorwart \cite{thorwart}, the interaction part
of our model Hamiltonian possesses symmetries (the Hamiltonian of the
two-qubit operation and the errors commute) that lead to better gate
quality factors. Furthermore, analysis of the symmetries and error
sources of our model system can lead to improved coupling schemes for
solid state qubits.
Milburn and co-workers on the other hand focused on comparison of classical and quantum
mechanical dynamics \cite{milburn} and estimated the decoherence
properties of two coupled two-state systems.

Governale \cite{governale} determined the decoherence properties of
two coupled charge qubits whose Hamiltonian differs from
(\ref{Hop2qb_nobath}) by the type of inter-qubit coupling, namely
$\sigma_y^{(1)} \otimes \sigma_y^{(2)}$ coupling.  However,
introducing the quality factors gives a measure to judge how certain
qubit designs perform quantum gate operations.

As a next step, one should consider driving, to be able to observe and
discuss Rabi oscillations in systems of two coupled qubits. It should
be investigated, how the decoherence properties are modified, if one
adds more qubits to the system.

We thank B.~Singh, L.~Tian, M.~Governale, M.C.~Goorden, A.C.J.\ ter~Haar,
H.~Aschauer, R.~Raussendorf, A.J.~Leggett, P.~H\"anggi, C.J.P.M.~Harmans, H.~Gutmann,
U.~Hartmann and J.~von Delft for fruitful discussions.

We acknowledge financial support from ARO, Contract-No. P-43385-PH-QC,
and DAAD, Contract-No.  DAAD/NSF D/0104619.

\end{document}